\documentclass[
aps,
prd,
amsmath,
floats,
floatfix,
twocolumn,
superscriptaddress,
nofootinbib,
showpacs,
10pt]{revtex4-1}
\usepackage{graphicx}
\usepackage{url}
\usepackage{amsmath,amssymb}
\usepackage{amsfonts}
\usepackage{bm}
\usepackage{xspace} % Sensible space treatment at end of simple macros
\usepackage[usenames]{color}
\usepackage{dcolumn}% Align table columns on decimal point
\usepackage[usenames]{color}

\usepackage{ulem}
\newcommand{\shift}{V}

\newcommand{\SMetric}{g}     % spatial metric
\newcommand{\SRicciS}{R}     % spatial Ricci scalar
\newcommand{\ExCurv}{K}      % spatial extrinsic curvature
\newcommand{\TrExCurv}{K}    % trace of spatial extrinsic curvature
\newcommand{\CF}{\psi}              % conformal factor

\newcommand{\CMetric}{{\tilde{g}}}    % conformal spatial metric
\newcommand{\Cornell}{\affiliation{Center for Radiophysics and Space
    Research, Cornell University, Ithaca, New York, 14853}}
\newcommand{\CITA}{\affiliation{Canadian Institute for Theoretical 
Astrophysics, University of Toronto, Toronto, Ontario M5S 3H8}}
\newcommand{\UNM}{\affiliation{Department of Mathematics and Statistics, 
The University of New Mexico, 
Albuquerque, New Mexico 87131}}
\begin{document}
\title{Implicit-explicit (IMEX) evolution of single black holes}
\author{Stephen R. Lau}\UNM 
\author{Geoffrey Lovelace}\Cornell
\author{Harald P. Pfeiffer}\CITA
\begin{abstract}
Numerical simulations of binary black holes---an 
important predictive tool for the detection of gravitational 
waves---are computationally expensive, especially for binaries 
with high mass ratios or with rapidly spinning constituent holes.
Existing codes for evolving binary black holes rely on explicit 
timestepping methods, for which the timestep size is limited by 
the smallest spatial scale through the Courant-Friedrichs-Lewy 
condition. Binary inspiral typically involves spatial scales (the 
spatial resolution required by a small or rapidly spinning hole) 
which are orders of magnitude smaller than the relevant (orbital, 
precession, and radiation-reaction) timescales characterizing 
the inspiral. Therefore, in explicit evolutions of binary black holes, 
the timestep size is typically orders of magnitude smaller than 
the relevant physical timescales. Implicit timestepping methods 
allow for larger timesteps, and they often reduce the 
total computational cost (without significant loss of accuracy) 
for problems dominated by spatial rather than temporal 
error, such as for binary-black-hole inspiral in corotating coordinates.
However, fully implicit methods can be difficult to implement for 
nonlinear evolution systems like the Einstein equations. Therefore, 
in this paper we explore implicit-explicit (IMEX) methods and 
use them for the first time to evolve black-hole spacetimes. 
Specifically, as a first step toward IMEX evolution of a full 
binary-black-hole spacetime, we develop an IMEX algorithm for the 
generalized harmonic formulation of the Einstein equations and use 
this algorithm to evolve stationary and perturbed single-black-hole 
spacetimes. Numerical experiments 
explore the stability and computational efficiency of our 
method.
\end{abstract}
\date{\today \hspace{0.2truecm}}
% 04.25.dg = numerical relativistic studies of black holes 
% 04.25.D- = numerical relativity
% 02.60.-x = numerical methods (mathematics)
% 02.70.-c = computational techniques; simulations
% 02.70.Jn = collocation methods 
\maketitle
%% \begin{center}
%% {\bf IMEX equations for the generalized harmonic system}\\[2mm]
%% {\sc Stephen R.~Lau}\\
%% Mathematics and Statistics\\
%% University of New Mexico\\
%% Albuquerque, NM 87131
%% \end{center}

%%%%%%%%%%%%%%%%%%%%%%%%%%%%%%%%%%%%%%%%%%%%%%%%%%%%%%%%%%%%%%%%
\section{Introduction}
%%%%%%%%%%%%%%%%%%%%%%%%%%%%%%%%%%%%%%%%%%%%%%%%%%%%%%%%%%%%%%%%

Binary black holes (BBHs) are important sources of gravitational waves for
the current and future gravitational wave detectors such as LIGO,
Virgo, LCGT~\cite{Barish:1999,Sigg:2008, Acernese:2008,Kuroda:2010}
and LISA~\cite{lisa_revised:2009, Jennrich:2009}.  Data-analysis of these
gravitational wave detectors proceeds with matched filtering, 
which requires accurate knowledge of the expected waveforms.  This
motivates numerical simulations of the inspiral, merger and ringdown
of two black holes.  Starting with Pretorius' 2005
breakthrough~\cite{Pretorius2005a}, several research groups have
developed numerical codes capable of simulating this process 
(see~\cite{Centrella:2010} for a recent review).

BBH inspiral simulations for gravitational wave detectors must cover 
at least the last $\approx 10$ orbits of the inspiral, and possibly 
many more~\cite{Santamaria:2010yb,Hannam:2010,Damour:2010,
MacDonald:2011ne,Boyle:2011dy}, requiring simulations significantly 
longer than the dynamical timescales of the individual black holes.  
This separation of temporal scales becomes particularly pronounced 
for a BBH with mass-ratio $q\gg 1$: The dynamical time of the smaller
black hole shrinks proportional to $1/q$.  Simultaneously, the
inspiral proceeds slower and the time the binary spends in the
strong-field regime lengthens proportionally to $q$.

All published numerical simulations of BBH inspiral and merger employ 
{\em explicit timestepping} algorithms which are subject to the 
Courant-Friedrichs-Lewy (CFL) condition which limits the 
timestep size by the smallest spatial scale in the problem. Binary 
inspiral typically involves spatial scales (the spatial resolution required 
by a small or rapidly spinning hole) which are orders of magnitude smaller 
than the relevant (orbital, precession, and radiation-reaction) 
timescales characterizing the inspiral. In explicit binary evolutions the 
CFL condition then effectively fixes the timestep size to be the dynamical 
timescale (see the last paragraph) for one of the constituent holes.
Such a timestep is orders of magnitude smaller than the
relevant physical timescales for the binary as a whole;
particularly when the binary has a large mass ratio (such as the 
simulations in Refs.~\cite{LoustoZlowchower:2011,Sperhake:2011ik})
or when at least one constituent hole has a high 
spin (since the horizon of the high-spin hole then requires higher 
spatial resolution).  For instance, a simulation with 
constituent holes with dimensionless spin magnitudes 
$0.95$~\cite{Lovelace2010} required half a million timesteps 
over 12.5 orbits. 

Were the CFL restriction overcome, computation of BBH inspirals 
with higher mass ratios, higher spins, and more orbits could 
become feasible. Implicit timestepping is one way to overcome 
the CFL condition and take larger timesteps. Of course, 
larger timesteps correspond to larger temporal truncation errors; 
however, a small timestep is required in BBH inspirals for 
{\em stability} (CFL condition) rather than {\em accuracy} 
(since, as argued above, the accuracy of a BBH inspiral 
is typically limited by spatial resolution, not temporal 
resolution). For problems dominated by spatial rather 
then temporal error, implicit timestepping methods often reduce 
the total computational cost (without significant
loss of accuracy), but fully implicit 
methods can be difficult to implement for nonlinear evolution 
systems like the Einstein equations. Implicit-explicit (IMEX) methods 
\cite{Dutt2000,Minion2003,LaytonMinion,HagstromZhou2006} are a 
compromise which we explore here. IMEX timestepping has been 
successfully applied to a variety of problems, including fluid-structure 
interaction~\cite{vanZuijlenEtAl:2007}, relativistic plasma 
astrophysics~\cite{PalenzuelaEtAl:2008}, and hydrodynamics with
heat conduction~\cite{KadiogluKnoll:2010}. 
In Ref.~\cite{LauPfeiffer2008}, Lau, Pfeiffer, and Hesthaven 
applied IMEX methods to evolve a forced 
scalar wave propagating on a curved spacetime (a Schwarzschild black 
hole), achieving stable evolutions with timestep sizes $\approx 1000$
times larger than with explicit methods.

In this paper, we lay much of the groundwork toward applying IMEX 
methods to full binary-black-hole evolutions. We develop an IMEX 
algorithm for one particular formulation of Einstein's equations
used in explicit BBH evolutions,
the generalized harmonic formulation (see \cite{Lindblom2006} and 
references therein). We use our IMEX algorithm 
to perform the first IMEX evolutions of single black holes 
(both static and dynamically perturbed). Our single-black-hole 
evolutions demonstrate the stability of our IMEX method. 
Further numerical experiments also investigate our method's efficiency; 
the IMEX algorithm offers a computational cost competitive with 
explicit evolution for sufficiently large step sizes. (Note that 
improved efficiency does not automatically follow from an IMEX 
algorithm affording larger timesteps, since each IMEX timestep 
is more expensive than an explicit step.) 
We also discuss further efficiency improvements of our IMEX 
implementation, and provide an outlook toward simulation of 
black hole binaries with IMEX techniques.

This paper is organized as follows. In Sec.~\ref{sec:math}, 
we derive the IMEX generalized harmonic equations and boundary 
conditions that we will use. In Sec.~\ref{sec:NumericalExperiments}, 
we explore numerical simulations using these equations, with a 
particular focus on the stability and efficiency gains of these 
simulations. We conclude in Sec.~\ref{sec:Discussion} by 
discussing the implications of our results, emphasizing 
the probable gains in computational efficiency when using 
IMEX in full binary-black-hole simulations.

%%%%%%%%%%%%%%%%%%%%%%%%%%%%%%%%%%%%%%%%%%%%%%%%%%%%%%%%%%%%%%%%
\section{IMEX formulation of 
         Einstein's equations}\label{sec:math}
%%%%%%%%%%%%%%%%%%%%%%%%%%%%%%%%%%%%%%%%%%%%%%%%%%%%%%%%%%%%%%%%

The generalized harmonic formulation of Einstein's equations
consists of ten coupled scalar wave equations.  Therefore, the present
discussion will borrow heavily from our earlier work on IMEX evolutions
of scalar fields on curved backgrounds~\cite{LauPfeiffer2008}.

%%%%%%%%%%%%%%%%%%%%%%%%%%%%%%%%%%%%%%%%%%%%%%%%%%%%%%%%%%%%%%%%
\subsection{Generalized harmonic system}
\label{sec:GH}
%%%%%%%%%%%%%%%%%%%%%%%%%%%%%%%%%%%%%%%%%%%%%%%%%%%%%%%%%%%%%%%%

Our goal is to solve Einstein's equations for the spacetime
metric $\psi_{ab}$, where Latin indices from the start of the
alphabet ($a, \ldots, f$) range over $0, 1, 2, 3$. The first order
{\em generalized harmonic} formulation of the Einstein evolution
equations given by Lindblom {\em et al} (Eqs.~(35)--(37) of
Ref.~\cite{Lindblom2006}) is the following:
\begin{subequations}\label{eq:GhSystem}
\begin{align}
\partial_t\psi_{ab} & =  
                      (1+\gamma_1)\shift^k \partial_k\psi_{ab}
                    - N \Pi_{ab} - \gamma_1 \shift^k \Phi_{kab}\\
\label{eq:GH-Pi}
\partial_t \Pi_{ab} & = 
                      \shift^k\partial_k \Pi_{ab} 
                    - N g^{jk} \partial_j \Phi_{kab}
                    + \gamma_1\gamma_2 \shift^k
                      \partial_k \psi_{ab} \nonumber \\
                        & 
                    + 2N \psi^{cd}\big(g^{jk}\Phi_{jca}\Phi_{kdb} 
                    - \Pi_{ca} \Pi_{db} 
                    - \psi^{ef}\Gamma_{ace}\Gamma_{bdf}
                      \big)\nonumber\\
                        & 
                    - 2N \nabla_{(a}H_{b)} 
                    - {\textstyle \frac{1}{2}}N t^c t^d \Pi_{cd}\Pi_{ab} 
                    - N t^c \Pi_{cj} g^{jk} \Phi_{kab} 
                        \nonumber \\
                        & 
                    + \gamma_0 N\big(2\delta^{c}{}_{(a} t_{b)} 
                    - \psi_{ab}t^c\big)\big(H_c+\Gamma_c\big) 
                    - \gamma_1 \gamma_2 \shift^k \Phi_{kab}\\
\partial_t \Phi_{jab} & = \shift^k \partial_k \Phi_{jab} 
                      - N \partial_j\Pi_{ab} 
                      + N \gamma_2\partial_j \psi_{ab}\nonumber \\
                        & 
                      + {\textstyle \frac{1}{2}} N t^c t^d \Phi_{jcd}\Pi_{ab} 
                      + N g^{km} t^c \Phi_{jkc}\Phi_{mab} 
                      - N\gamma_2 \Phi_{jab}.
\end{align}
\end{subequations}
Here, $N$, $V^k$, and $g_{jk}$ are the spacetime metric's 
associated lapse function, shift vector, and spatial metric 
induced on level-$t$ slices. Latin indices from the middle 
of the alphabet $i, j, \ldots=1,2,3$ range only over spatial 
dimensions. As a one-form, $t_a = -N \partial_a t$ is the 
unit normal to the temporal foliation defined by the coordinate 
time $t$. The other fundamental variables 
$\Pi_{ab} \equiv -t^c \partial_c\psi_{ab}$ and $\Phi_{kab} \equiv
\partial_k\psi_{ab}$ arise from the reduction of the 
generalized harmonic equations to first order form. The latter 
definition leads to the auxiliary constraint
\begin{equation}\label{eq:3indexC}
\mathcal{C}_{kab}
\equiv \partial_k \psi_{ab} - \Phi_{kab}=0.
\end{equation}
The variable $\Gamma_a = \psi^{bc}\Gamma_{abc}$ 
represents a contraction of the Christoffel symbols 
$\Gamma_{abc}$ of the spacetime metric
$\psi_{ab}$. Time 
derivatives $\partial_t\psi_{ab}$ 
inside $\Gamma_{abc}$ are evaluated in terms of 
$N$, $V^k$, $\Pi_{ab}$, and $\Phi_{kab}$~\cite{Lindblom2006}.

The functions $H_c$ are freely specifiable and embody the
coordinate-freedom of Einstein's equations~\cite{Lindblom2006}.  
Einstein's equations can be written as a set of 
constrained evolution equations; in the 
generalized harmonic formulation, the fundamental 
constraint takes the form 
\begin{equation}\label{eq:1indexC}
\mathcal{C}_a \equiv H_a + \Gamma_a=0.
\end{equation}
Constraint damping~\cite{Gundlach2005,Pretorius2005a,Lindblom2006,Holst2004} 
is used to enforce both the fundamental constraint~(\ref{eq:1indexC})
and the auxiliary constraint~(\ref{eq:3indexC}). Those terms
in Eqs.~(\ref{eq:GhSystem}) proportional to $\gamma_0$ damp the 
fundamental constraint~(\ref{eq:1indexC}). 
Those terms proportional to $\gamma_{1}$ and $\gamma_{2}$ 
in Eqs.~(\ref{eq:GhSystem}) damp the constraint \eqref{eq:3indexC}.
Our IMEX formulation converts to second order variables and so the 
auxiliary constraint is trivially satisfied.  Therefore, in the 
rest of this paper, we set $\gamma_1 = 0 = \gamma_2$ in all IMEX 
evolutions. 

%%%%%%%%%%%%%%%%%%%%%%%%%%%%%%%%%%%%%%%%%%%%%%%%%%%%%%%%%%%%%%%%
\subsection{First-order implicit equations and second-order
implicit equation for the metric}
\label{sec:ImexFoshSplit}
%%%%%%%%%%%%%%%%%%%%%%%%%%%%%%%%%%%%%%%%%%%%%%%%%%%%%%%%%%%%%%%%

Although \eqref{eq:GhSystem} is a system of partial 
differential equations (PDEs), 
we formally view it as an ordinary differential equation (ODE) 
initial value problem,
\begin{equation}\label{eq:IVP}
\frac{d\boldsymbol{u}}{dt} =
\boldsymbol{f}(t,\boldsymbol{u}),
\quad
\boldsymbol{u}(t_0) = \boldsymbol{u}_0,
\end{equation}
so that our notation conforms with the literature 
\cite{Dutt2000,Minion2003,LaytonMinion,HagstromZhou2006}
on IMEX ODE methods. [Otherwise, we would have used 
partial time differentiation in \eqref{eq:IVP}.]
The system \eqref{eq:GhSystem} is actually also solved 
as an initial boundary value problem; however, we defer 
the issue of boundary conditions to a later subsection.
In this view $\boldsymbol{u}$ represents the 
collection
$(\psi_{ab},\Pi_{ab},\Phi_{kab})$ of fundamental
fields. Furthermore, we assume there exists a splitting
\begin{equation}
\boldsymbol{f}(t,\boldsymbol{u})   =
\boldsymbol{f}^I(t,\boldsymbol{u}) +
\boldsymbol{f}^E(t,\boldsymbol{u})
\end{equation}
of the right-hand side $\boldsymbol{f}$ into an explicit
sector $\boldsymbol{f}^E$ and an implicit sector $\boldsymbol{f}^I$.
In this paper, as in Ref.~\cite{LauPfeiffer2008}, we {\em
split by equation}. That is, we choose which terms on the
right-hand side of Eq.~\eqref{eq:GhSystem} are to be treated
implicitly. 

To take a timestep, we choose an IMEX timestepping algorithm,
such as ImexEuler, 
Additive Runge Kutta (ARK)
\cite{Kennedy-Carpenter:2003}, 
or semi-implicit spectral-deferred correction (SISDC) 
\cite{Dutt2000,Minion2003,LaytonMinion,HagstromZhou2006}.
We note that while ARK was used almost exclusively in
Ref.~\cite{LauPfeiffer2008}, we have encountered stability
issues with its use in the work presented here, and therefore
focus here on SISDC. As explained in Sec.~II A of 
Ref.~\cite{LauPfeiffer2008}, each of these algorithms requires 
that we are able to solve (multiple times per timestep) an 
implicit equation of the form
\begin{equation}
\boldsymbol{u} - \alpha\boldsymbol{f}^I(t,\boldsymbol{u})
= \boldsymbol{B},
\end{equation}
where $\alpha$ is proportional to the step size $\Delta t$
and the inhomogeneity $\boldsymbol{B}$ is defined by the
algorithm. For example, the corresponding equation for 
ImexEuler integration,
\begin{equation}
\boldsymbol{u}_{n+1} - 
\Delta t\boldsymbol{f}^I(t_{n+1},\boldsymbol{u}_{n+1})
= \boldsymbol{u}_n     + 
\Delta t\boldsymbol{f}^E(t_n,\boldsymbol{u}_n),
\end{equation}
is solved to advance the solution from time $t_n$ to time
$t_{n+1}$. Concrete expressions for $\boldsymbol{B}$ are
given in Ref.~\cite{LauPfeiffer2008} for ARK and in 
Appendix~\ref{sec:SISDC} for SISDC. 

The IMEX splitting of the system \eqref{eq:GhSystem} 
that we chose is analogous to the ``case (ii)" equations 
for the scalar-wave system given as Eqs.~(15a)--(15c) in 
Ref.~\cite{LauPfeiffer2008}.  Specifically, we treat implicitly 
the entire right-hand sides of Eqs.~(\ref{eq:GhSystem}a) and 
(\ref{eq:GhSystem}c). However, a fully implicit treatment of 
the equation for $\Pi_{ab}$ has turned out 
to be prohibitively complicated. Therefore, 
of the terms appearing in the right-hand side of 
Eq.~(\ref{eq:GhSystem}b), we have chosen to include in the 
implicit sector only the principal-part terms and, possibly, 
the constraint damping term proportional to $\gamma_0$. 
The principal-part terms are the stiff terms which 
most constrain the timestep size, and, as we shall see later,
the constraint damping term is also stiff. Implicit treatment 
of the remaining terms on the right-hand side of 
Eq.~(\ref{eq:GhSystem}b) would be difficult because the 
implicit equation which results from their inclusion 
has an extremely complicated variation. This variation 
would be required were the resulting equation solved 
(as part of the overall system) via 
Newton iteration.

Our splitting of Eq.~(\ref{eq:GhSystem}b) could be improved upon. 
Indeed, with $\boldsymbol{f}_{\Pi_{ab}}(t,\boldsymbol{u})$ 
representing the right-hand side of the evolution equation 
(\ref{eq:GhSystem}b) for $\Pi_{ab}$, a binary evolution based on 
the dual-frames approach will have $\boldsymbol{f}_{\Pi_{ab}} = 
\mathcal{O}(\omega)$, where $\omega$ is the orbital frequency
(a small quantity). However, for our described splitting 
both $\boldsymbol{f}^I_{\Pi_{ab}}$ and 
$\boldsymbol{f}^E_{\Pi_{ab}}$ would be $\mathcal{O}(1)$. 
Although their combination is small, each individual term
on the right-hand side of (\ref{eq:GhSystem}b) need not be. 
In other words, there appears to be no natural {\em splitting 
by equation} for Eqs.~(\ref{eq:GhSystem}), as there often is 
for, say, advection-diffusion problems. While we do not yet 
fully appreciate the
consequences of the splitting we shall
employ here, we are considering approaches to mitigate
potential problems with our splitting-by-equation approach.
Among these is a fully implicit implementation of 
Eq.~(\ref{eq:GhSystem}b), with other possibilities discussed
in the conclusion of Ref.~\cite{LauPfeiffer2008}.

Our choices above correspond to the following 
first-order implicit equation for $\Pi_{ab}$:
\begin{align}\label{eq:ImpPiequation}
\begin{split}
\Pi_{ab} & - \alpha \big[\shift^k\partial_k \Pi_{ab} 
         - N g^{jk}\partial_j\Phi_{kab}\\
         & + \gamma_0^I N\big(2\delta^{c}{}_{(a} t_{b)}
                    - \psi_{ab}t^c\big)\big(H_c+\Gamma_c\big)\big]
= B_{\Pi_{ab}}.
\end{split}
\end{align}
Here we have split the damping parameter as
$\gamma_0 = \gamma_0^I + \gamma_0^E$, which in general
allows for part of the damping term to be treated
implicitly (if $\gamma_0^I \neq 0$) and part explicitly
(if $\gamma_0^E \neq 0$). In Eq.~\eqref{eq:ImpPiequation}
we view $\Gamma_e$ as 
\begin{equation}\label{eq:decompGamma}
\Gamma_e = \underbrace{\frac{\partial \Gamma_e}{
           \partial\Pi_{cd}} \Pi_{cd}}_{\text{terms with }\Pi_{ab}}
+ \underbrace{\left[\Gamma_e - \frac{
           \partial \Gamma_e}{\Pi_{cd}} 
           \Pi_{cd}\right]}_{\text{terms without }\Pi_{ab}},
\end{equation}
with the details of this decomposition given in 
Appendix~\ref{sec:GammaDecomp}. The reason for the decomposition 
is given immediately after Eq.~\eqref{eq:alfaNPi}. In all, our 
first--order implicit equations corresponding to 
the evolution system \eqref{eq:GhSystem} 
are then as follows:
\begin{subequations}\label{eq:ImexGhCase2}
\begin{align}
\psi_{ab}  & - \alpha\big(\shift^k \partial_k\psi_{ab}
- N \Pi_{ab}\big) = B_{\psi_{ab}}\\
\Pi_{ab}   & - \alpha \big(
\shift^k\partial_k \Pi_{ab} - N g^{jk} \partial_j \Phi_{kab}
+ N\mathcal{Q}_{ab}{}^{cd}\Pi_{cd} 
\nonumber \\
& + N \mathcal{G}_{ab}
\big) = B_{\Pi_{ab}}\\
\Phi_{jab} & - \alpha\big(\shift^k \partial_k \Phi_{jab}
- N \partial_j\Pi_{ab} 
+ {\textstyle \frac{1}{2}} N t^c t^d \Phi_{jcd}\Pi_{ab}
\nonumber \\
           & + N g^{km} t^c \Phi_{jkc}\Phi_{mab}\big) 
             = B_{\Phi_{jab}},
\end{align}
\end{subequations}
where
\begin{align}\label{eq:QandG}
\mathcal{Q}_{ab}{}^{cd}  
& \equiv \gamma_0^I \big(2\delta^{e}{}_{(a} t_{b)}
- \psi_{ab}t^e\big)\frac{\partial\Gamma_e}{\partial \Pi_{cd}}
\nonumber \\
\mathcal{G}_{ab}  
& \equiv \gamma_0^I \big(2\delta^{e}{}_{(a} t_{b)}
- \psi_{ab}t^e\big)\left[H_e+\Gamma_e
- \frac{\partial\Gamma_e}{\partial \Pi_{cd}}\Pi_{cd}\right].
\end{align}

To solve these equations, we first take a combination of 
them to get a single second-order equation for $\psi_{ab}$.
In terms of $\xi_{ab} \equiv \psi_{ab} - \alpha \shift^k 
\partial_k\psi_{ab}$, we express (\ref{eq:ImexGhCase2}a) as
\begin{equation}\label{eq:xifirsteqn}
\alpha N \Pi_{ab} = B_{\psi_{ab}} - \xi_{ab}.
\end{equation}
Multiplication of Eq.~(\ref{eq:ImexGhCase2}b) by $\alpha N$,
followed by a substitution with \eqref{eq:xifirsteqn},
yields
\begin{align}\label{eq:alfaNPi}
   & \alpha N \Pi_{ab} 
   - \alpha^2 N\shift^k\partial_k \Pi_{ab} 
   + \alpha^2 N^2 g^{jk} 
     \partial_j \Phi_{kab} 
\nonumber \\
   & - \alpha N \mathcal{Q}_{ab}{}^{cd}(B_{\psi_{cd}} - \xi_{cd})
   - \alpha^2 N^2 \mathcal{G}_{ab}
   = \alpha N B_{\Pi_{ab}}.
\end{align}
The decomposition (\ref{eq:decompGamma}) 
ensures that the substitution with Eq.~\eqref{eq:xifirsteqn} 
is also made for the $\Pi_{cd}$ terms in $\Gamma_e$.
We subtract the last equation from (\ref{eq:ImexGhCase2}a)
to reach
\begin{align}\label{eq:Piresult}
& \psi_{ab} - \alpha \shift^k \partial_k \psi_{ab}
+ \alpha^2 N\shift^k\partial_k \Pi_{ab} 
- \alpha^2 N^2 g^{jk} \partial_j \Phi_{kab}
\nonumber \\ 
& - \alpha N \mathcal{Q}_{ab}{}^{cd}\xi_{cd}
  + \alpha^2 N^2 \mathcal{G}_{ab}
\nonumber \\
& = B_{\psi_{ab}}
  - \alpha N B_{\Pi_{ab}}
  -\alpha N \mathcal{Q}_{ab}{}^{cd}B_{\psi_{cd}}.
\end{align} 
We must eliminate the term $\alpha^2 N 
\shift^k \partial_k\Pi_{ab}$ from the result. 
To this end, we contract 
Eq.~(\ref{eq:ImexGhCase2}c) into
$\alpha \shift^j$, thereby finding
\begin{align}
& \alpha \shift^j \Phi_{jab} -
\alpha^2 \shift^k \shift^j \partial_k \Phi_{jab}
+ \alpha^2 N \shift^j \partial_j\Pi_{ab}
\nonumber \\
& - {\textstyle \frac{1}{2}} \alpha^2 N t^c t^d 
  \shift^j \Phi_{jcd}\Pi_{ab} 
- \alpha^2 N g^{km} t^c \shift^j \Phi_{jkc}\Phi_{mab}
\nonumber \\
& = \alpha \shift^j B_{\Phi_{jab}},
\end{align}
which, using Eq.~(\ref{eq:xifirsteqn}), we rewrite as
\begin{align}
& \alpha \shift^j \Phi_{jab} - \alpha^2 \shift^k \shift^j 
  \partial_k \Phi_{jab}
+ \alpha^2 N \shift^j \partial_j\Pi_{ab}
\nonumber \\
& + {\textstyle \frac{1}{2}} \alpha t^c t^d \shift^j 
\Phi_{jcd} \xi_{ab}
- \alpha^2 N g^{km} t^c \shift^j \Phi_{jkc}\Phi_{mab}
\nonumber \\
& = \alpha \shift^j B_{\Phi_{jab}}
+ {\textstyle \frac{1}{2}} \alpha t^c t^d \shift^j
\Phi_{jcd} B_{\psi_{ab}}.
\end{align}
Subtracting the last equation from \eqref{eq:Piresult} and 
making substitutions with the constraint \eqref{eq:3indexC},
we arrive at the following second--order equation:
\begin{align}
\psi_{ab} & - 2\alpha \shift^k \partial_k \psi_{ab}
- \alpha^2 \big(N^2 g^{jk} - \shift^j \shift^k\big) 
\partial_j\partial_k \psi_{ab}
\nonumber \\
& - {\textstyle \frac{1}{2}} \alpha t^c t^d \shift^j 
  (\partial_j \psi_{cd})
  (\psi_{ab} - \alpha \shift^k \partial_k \psi_{ab})
\nonumber \\
& + \alpha^2 N g^{km} t^c \shift^j 
  (\partial_j \psi_{kc})(\partial_m \psi_{ab})
\nonumber \\
& - \alpha N \mathcal{Q}_{ab}{}^{cd}
    (\psi_{cd} - \alpha V^k \partial_k \psi_{cd})
  + \alpha^2 N^2 \mathcal{G}_{ab}
\nonumber \\
& = \big(1-{\textstyle \frac{1}{2}} \alpha t^c t^d \shift^j
\partial_j \psi_{cd}\big)B_{\psi_{ab}} 
- \alpha N B_{\Pi_{ab}} - \alpha \shift^k B_{\Phi_{kab}}
\nonumber \\
& -\alpha N \mathcal{Q}_{ab}{}^{cd}B_{\psi_{cd}}
+ \text{ terms homogeneous in } \mathcal{C}_{kab}.
\label{eq:psiEquation}
\end{align}
To solve the system \eqref{eq:ImexGhCase2}, we first 
solve \eqref{eq:psiEquation}, subject to boundary conditions 
discussed in Sec.~\ref{sec:BC}. Next, we recover $\Pi_{ab}$ 
algebraically from (\ref{eq:ImexGhCase2}a). Finally, we  
set $\Phi_{kab} = \partial_k \psi_{ab}$, i.e., we 
enforce that the constraint $\mathcal{C}_{kab} = 0$. 

We stress that, as a linear and undifferentiated combination of 
Eqs.~\eqref{eq:ImexGhCase2} for the first-order system, 
Eq.~\eqref{eq:psiEquation} actually contains no second-order 
derivatives of $\psi_{ab}$. Indeed, all of the $B$-terms
on the right-hand side of Eq.~\eqref{eq:psiEquation} appear 
undifferentiated, indicating that we have not differentiated
the first-order system \eqref{eq:ImexGhCase2}. Each second-order 
derivative of $\psi_{ab}$ on the left-hand side of 
\eqref{eq:psiEquation} is precisely canceled by a corresponding 
term appearing in one of the constraint terms on the 
right-hand side [not shown explicitly in 
Eq.~(\ref{eq:psiEquation})]. Now, 
when \emph{numerically} solving  Eq.~\eqref{eq:psiEquation}, we 
set the constraint terms from the right-hand side to zero, 
thereby creating a genuinely second-order equation. We discuss 
the permissibility of this procedure in Sec.~\ref{sec:AuxCon} below.

\subsection{Boundary conditions}\label{sec:BC}
For 
black-hole evolutions which employ excision, the inner
boundary lies within an apparent horizon. For this 
scenario we adopt no inner boundary condition, regardless
of what condition is adopted at the outer boundary and despite the fact
that Eq.~\eqref{eq:psiEquation} is a second-order equation.
In the context of scalar fields on a fixed black-hole
background, Ref.~\cite{LauPfeiffer2008} has discussed
the motivation for and permissibility of this procedure.
A similar analytical treatment of the coupled 
nonlinear system \eqref{eq:psiEquation} would be, we
suspect, a difficult piece of mathematical analysis, 
one beyond the scope of this paper. Therefore, here 
we content ourselves both with the scalar field analogy 
and the observation that the lack of an inner boundary 
condition has caused no difficulties numerically. 
Nevertheless, the issue merits further study.

The outer boundary condition that we apply to 
Eq.~\eqref{eq:psiEquation} is either (i) a fixed Dirichlet condition 
on each component $\psi_{ab}$ of the spacetime metric or 
(ii) the following condition. In terms of the incoming characteristic 
variable $U^{-}_{ab} \equiv \Pi_{ab} - n^k \Phi_{kab}$ 
(where $n^k$ is the unit, outward-pointing, 
normal vector to the boundary), we rewrite 
Eq.~(\ref{eq:ImexGhCase2}a) as
\begin{equation}\label{eq:ImexFixUminus}
\psi_{ab} + \alpha (N n^k - \shift^k) \partial_k\psi_{ab} = 
B_{\psi_{ab}} - \alpha N U^{-}_{ab} + \alpha N n^k 
\mathcal{C}_{kab},
\end{equation}
We control $U^{-}_{ab}$
at the boundary; therefore, both $B_{\psi_{ab}}$ and
$U^{-}_{ab}$ here appear as fixed quantities, and 
Eq.~(\ref{eq:ImexFixUminus}) represents a boundary 
condition on $\psi_{ab}$. Moreover,
when numerically enforcing this condition we also set
the constraint term on the right-hand side to zero. 

\subsection{Implicit equation for the auxiliary constraint}
\label{sec:AuxCon}
Eqs.~(\ref{eq:ImexGhCase2}a) and (\ref{eq:ImexGhCase2}c) 
imply an implicit equation
for the auxiliary constraint. Partial differentiation 
of (\ref{eq:ImexGhCase2}a) yields
\begin{align}
\partial_j\psi_{ab} 
& - \alpha\big[ 
(\partial_j\shift^k) 
(\partial_k\psi_{ab}) 
+ \shift^k \partial_k\partial_j \psi_{ab}
\nonumber \\
& 
- (\partial_j N) \Pi_{ab}
 - N \partial_j\Pi_{ab}\big) 
= \partial_j B_{\psi_{ab}}.
\label{eq:psiIMPdiff}
\end{align}
To express the derivatives of the lapse and shift in terms
of derivatives of the metric $\psi_{ab}$, we use the result 
\begin{equation}
\delta\psi_{ab} = -2N^{-1}t_a t_b \delta N
                  -2N^{-1} g_{k(a} t_{b)} \delta\shift^k
                  +g^i_{(a} g^k_{b)} \delta g_{ik},
\end{equation}
which in turn yields
\begin{equation}
\delta N = -{\textstyle \frac{1}{2}}
           N t^c t^d \delta\psi_{cd},\quad
\delta V^k = N g^{km} t^c \delta\psi_{mc}.
\end{equation}
Insertion of these results (with the variation
$\delta \rightarrow \partial_j$) into \eqref{eq:psiIMPdiff} gives
\begin{align}
\partial_j\psi_{ab}    
& - \alpha\big[
N g^{km} t^c (\partial_j\psi_{mc})(\partial_k\psi_{ab}) 
+ \shift^k \partial_k\partial_j \psi_{ab}
\nonumber \\
& 
+{\textstyle \frac{1}{2}}N t^c t^d (\partial_j\psi_{cd})
\Pi_{ab}
- N \partial_j\Pi_{ab}\big] = \partial_j B_{\psi_{ab}}.
\end{align}
Finally, we subtract (\ref{eq:ImexGhCase2}c) from the 
last equation and make substitutions with the constraint
to reach 
\begin{align}\label{eq:ImpEquationC}
\mathcal{C}_{jab} &
                    - \alpha\big[
                      \shift^k \partial_k \mathcal{C}_{jab}
                    + N g^{km} t^c
                      (\Phi_{jkc}\mathcal{C}_{mab}
                      +\mathcal{C}_{jkc}\partial_m\psi_{ab})
\nonumber \\
&
                    +{\textstyle \frac{1}{2}}
                     N t^c t^d \mathcal{C}_{jcd}\Pi_{ab}\big]
                    =\partial_j B_{\psi_{ab}} - B_{\Phi_{jab}}.
\end{align}

This equation is analogous to Eq.~(20) of Ref.~\cite{LauPfeiffer2008},
\begin{equation}
\bar{\mathcal{C}}_j - \alpha \pounds_V \bar{\mathcal{C}}_j = 
\partial_j B_\psi - B_{\Phi_j},
\label{eq:scalarimpeq}
\end{equation}
for scalar waves on a fixed curved background, where 
the overbar on $\bar{\mathcal{C}}_j$ serves to differentiate
this constraint from the generalized harmonic constraint 
$\mathcal{C}_a$ in Eq.~\eqref{eq:1indexC} (which carries a 
spacetime rather than spatial index in any case). Specifically, 
in the scalar wave scenario the variables $(\psi,\Pi,\Phi_k)$ 
are analogous to the generalized harmonic variables 
$(\psi_{ab},\Pi_{ab},\Phi_{kab})$, and the auxiliary constraint 
is $\bar{\mathcal{C}}_j \equiv \partial_j\psi - \Phi_j$.
Starting with a prescribed $\bar{\mathcal{C}}_j$ at the outer boundary, 
we may integrate Eq.~(\ref{eq:scalarimpeq})
along the integral curves of the shift vector. This independent 
integration of $\bar{\mathcal{C}}_j$ proved important toward understanding 
in what sense solving the second-order implicit equation for $\psi$
[analogous to Eq.~\eqref{eq:psiEquation}] was equivalent to solving the 
first-order system for $(\psi,\Pi,\Phi_k)$ [analogous to 
Eq.~\eqref{eq:ImexGhCase2}].
Such an independent integration of \eqref{eq:ImpEquationC} is 
clearly not possible. Nevertheless, provided both $\mathcal{C}_{jab} = 0$ 
on the outer boundary and a vanishing right-hand source in 
\eqref{eq:ImpEquationC}, the equation formally determines 
$\mathcal{C}_{jab} = 0$ along the integral curves of $V^k$.
This motivates our neglecting the terms homogeneous in 
$\mathcal{C}_{jab}$ in Eq.~(\ref{eq:psiEquation}).

Consideration of our steps above for solving
  \eqref{eq:ImexGhCase2} shows that the constraint $\mathcal{C}_{jab}$
  remains exactly zero throughout our IMEX scheme. We are then
  effectively evolving only the variables $\psi_{ab}$ and
  $\Pi_{ab}$. Our reasons for nevertheless retaining $\Phi_{jab}$ in
  the formalism are twofold. First, {\tt SpEC} ---the software project we
  have used for simulations--- chiefly supports first order symmetric 
  hyperbolic systems.
  Second, as described in the conclusion, for the binary problem we
  envision a {\em split by region} approach, in which outer subdomains
  are treated explicitly and inner subdomains (spherical shells)
  immediately near the holes are treated by IMEX methods. Since
  explicit evolutions in {\tt SpEC} currently require a
  first order system, the variable $\Phi_{jab}$ must be present in
  the outer subdomains. Coupling between the outer and inner
  subdomains is then facilitated by having $\Phi_{jab}$ also available 
  on the inner subdomains. There has been recent progress in
  applying spectral methods to evolve second order in space partial
  differential equations~\cite{Taylor:2010ki}.  If these techniques
  work for the generalized harmonic system, 
  it should be possible to abandon $\Phi_{jab}$ entirely.

%%%%%%%%%%%%%%%%%%%%%%%%%%%%%%%%%%%%%%%%%%%%%%%%%%%%%%%%%%%%%%%%
\section{Numerical Experiments}
\label{sec:NumericalExperiments}
%%%%%%%%%%%%%%%%%%%%%%%%%%%%%%%%%%%%%%%%%%%%%%%%%%%%%%%%%%%%%%%%
Through numerical simulations of single black holes, we 
now examine the behavior of the scheme presented above.  We evolve 
initial data representing both (i) the static Schwarzschild solution
in Kerr-Schild coordinates and (ii) the same solution with a superposed 
ingoing pulse of gravitational radiation. The latter is a vacuum problem 
with non-trivial evolution. As the gravitational wave pulse 
travels inward, it hits and perturbs the black hole.  Most of 
the pulse is absorbed by the black hole, increasing its mass; 
the rest is scattered and propagates away. This test features 
initial dynamics on short timescales (moving pulse of radiation, 
perturbed black hole), 
with relaxation to time-independence. Eventually, the black hole 
settles down to a stationary black hole, and the scattered radiation 
leaves the computational domain through the outer 
boundary. Technical details for the dynamical case (ii)
are summarized in Appendix~\ref{App:ID}.

\subsection{Long-time stability of IMEX evolutions} \label{sec:stability}
In this subsection we demonstrate the stability of our IMEX 
algorithm by evolving the static Schwarzschild solution in 
Kerr-Schild coordinates to late times (up to $10^4M$), 
adopting fixed Dirichlet conditions, that is with $\psi_{ab}$ 
fixed as the analytical solution on the outer boundary. We 
note that the radiation conditions (\ref{eq:ImexFixUminus}), 
with $U^{-}_{ab}$ determined by the analytical solution on 
the outer boundary, apparently give rise to an extremely weak 
instability. Indeed, with Eq.~(\ref{eq:ImexFixUminus}) a slowly 
growing instability appears after (sometimes well after) time 
$10^3 M$. We specify no inner boundary condition 
(cf. Sec.~\ref{sec:BC}). Our domain, 
a single spherical shell with Cartesian center
$(0.01,-0.0097,0.003)$, is determined by a top spherical 
harmonic index $\ell_\mathrm{max} = 7$ and the radial interval 
$1.9 \leq r \leq 11.9$, with $N_r = 15$ radial collocation 
points and an exponential mapping of the radial coordinate 
(see Eq.~(48) of \cite{LauPfeiffer2008}). Results 
for Cartesian center $(0,0,0)$ are qualitatively similar, but 
with the corresponding errors a few orders of magnitude smaller.
For constraint damping parameters, we 
have taken $\gamma_0^I = 1$ and $\gamma_0^E = 0$. 

\begin{figure}
\includegraphics[width=0.45\textwidth]{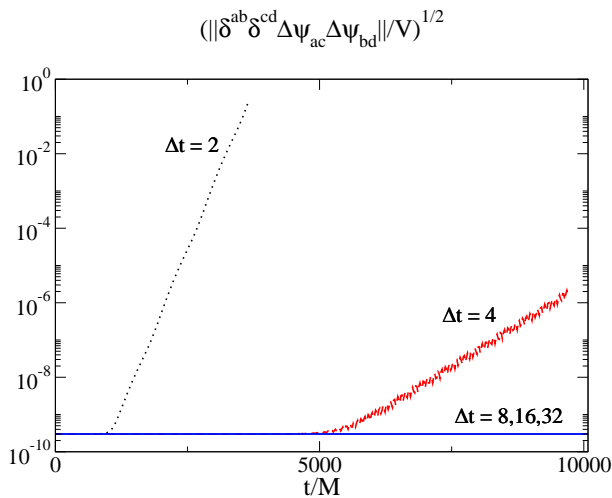}
\caption{Error histories for GLoSISDC3.
$\|\cdot\|$ represents the 1-norm with respect to the 
Cartesian coordinate measure over the spherical shell,
i.~e.~$\|f\| = \int_V |f|dxdydz$, and $V$
is the improper (coordinate) volume of the spherical shell. 
As mentioned in the text, $\Delta\psi_{ab}$ 
denotes the difference between the numerical
metric and the exact solution.
}
\label{fig:LongTimeErrorsGLo}
\end{figure}

\begin{figure}
\includegraphics[width=0.45\textwidth]{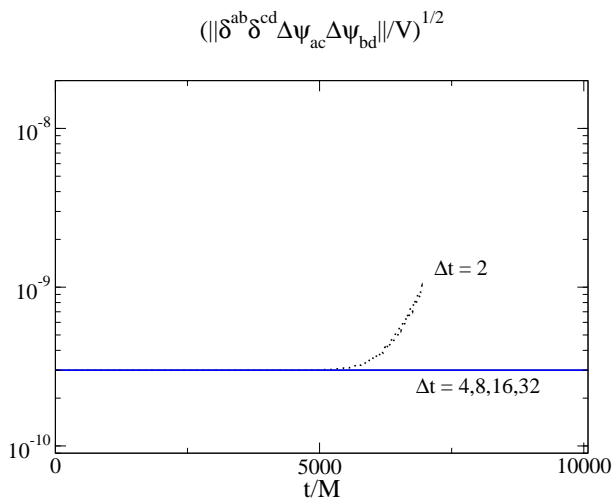}
\caption{Error histories for GRrSISDC2. See the 
caption of Figure~\ref{fig:LongTimeErrorsGLo} for an 
explanation of the figure labels.
}
\label{fig:LongTimeErrorsGRr}
\end{figure}

We have performed IMEX evolutions with an ImexEuler 
timestepper (first order accurate and requiring one solution 
of the system (\ref{eq:ImexGhCase2}) per timestep), 
3-point (substep) Gauss-Lobatto SISDC (GLoSISDC3, fourth 
order accurate, eight implicit solves per timestep), and 
2-point (substep) Gauss-Radau-right SISDC (GRrSISDC2, 
third order accurate, six implicit solves per timestep). 
Since the geometry is time-independent, numerical solution 
of \eqref{eq:psiEquation} will be achieved 
without any iterations in the Newton-Raphson algorithm, 
assuming that the solution at the previous 
timestep serves as an initial guess. To prevent this trivial 
convergence, we have rescaled the initial guess $\psi^0_{ab}
\rightarrow 1.00001 \psi^0_{ab}$ before each implicit solve. 
For GLoSISDC3 and GRrSISDC2 respectively, 
Figs.~\ref{fig:LongTimeErrorsGLo} and 
\ref{fig:LongTimeErrorsGRr} depict error histories for the metric 
$\psi_{ab}$ as measured against the exact solution. Each plot
exhibits long-time stability for the larger timesteps considered
but weak instability for some of the smaller timesteps.
%%%%%%%%%%%%%%%%%%%%%%%%%%%%%%%%%%%%%%%%%%%%%%%%%%%%%%%%%%%%%%%%
%%%%%%%%%%%%%%%%%%%%%%%%%%%%%%%%%%%%%%%%%%%%%%%%%%%%%%%%%%%%%%%%
%%%%%%%%%%%%%%%%%%%%%%%%%%%%%%%%%%%%%%%%%%%%%%%%%%%%%%%%%%%%%%%%
%%%%%%%%%%%%%%%%%%%%%%%%%%%%%%%%%%%%%%%%%%%%%%%%%%%%%%%%%%%%%%%%

\begin{figure}
\includegraphics[width=0.45\textwidth]{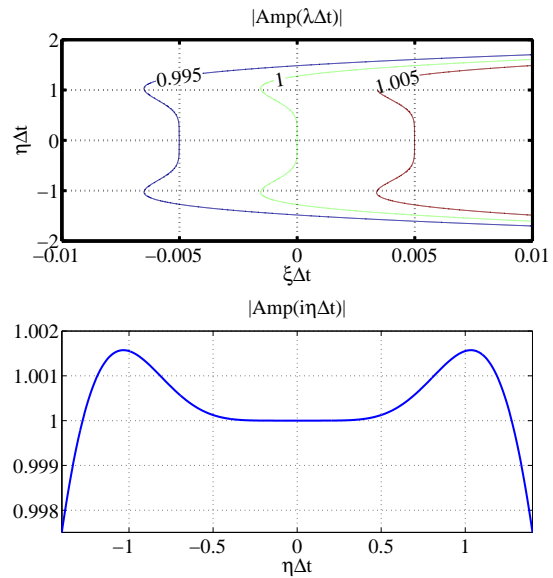}
\caption{Diagram for implicit sector of GLoSISDC3.
The bottom
plot depicts the cross section of the top plot along the imaginary
axis, with $\lambda = \mathrm{i}\eta \in \mathrm{i}\mathbb{R}$.
}
\label{fig:ImpGLoStabilityDiagram}
\end{figure}

\begin{figure}
\includegraphics[width=0.45\textwidth]{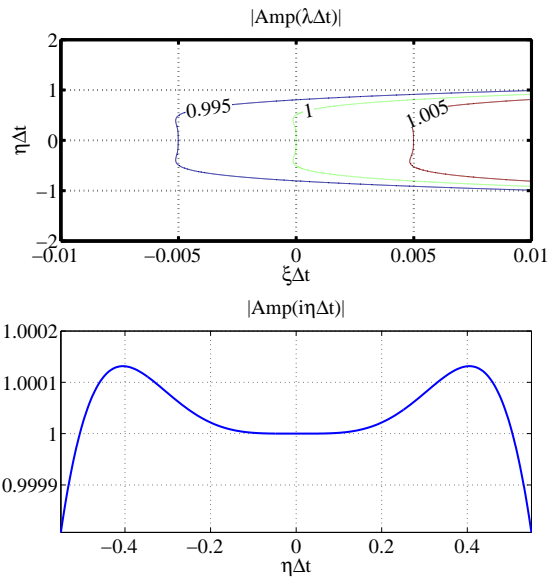}
\caption{Diagram for implicit sector of GRrSISDC2. See 
relevant comments given in the caption of 
Fig.~\ref{fig:ImpGLoStabilityDiagram}. 
}
\label{fig:ImpGRrStabilityDiagram}
\end{figure}

Examination of the stability diagrams for these methods suggests
a heuristic explanation of our results. The diagram for a given
(either explicit or implicit) ODE method is determined by 
its application to the model problem $du/dt = \lambda u$, where 
$\lambda = \xi + \mathrm{i}\eta$. 
Subject to the initial condition $u_0 = 1$, a single timestep 
for a given method produces an update $u_{\Delta t} = 
\mathrm{Amp}(\lambda \Delta t)$, the {\em amplification factor}
which is a function of the complex variable 
$\lambda \Delta t$. The {\em region of absolute stability} for 
a given method is then the domain in the $(\lambda \Delta t)$-plane 
for which $|\mathrm{Amp}(\lambda \Delta t)| \leq 1$. Figures 
\ref{fig:ImpGLoStabilityDiagram} and \ref{fig:ImpGRrStabilityDiagram}
respectively depict the stability diagrams for GLoSISDC3 and 
GRrSISDC2, with the model problem treated fully 
implicitly, i.e.~with $f^I = \lambda u$ and $f^E = 0$. 
For both diagrams, our interest lies with the imaginary 
axis, since the system \eqref{eq:GhSystem}
of equations we evolve supports the propagation of waves.

For GLoSISDC3, the imaginary axis lies within the region of 
absolute stability, except for a portion around the origin.
The bottom panel of 
Fig.~\ref{fig:ImpGLoStabilityDiagram} shows that 
$|\mathrm{Amp}(\mathrm{i}\eta\Delta t)|>1$ for 
$|\eta\Delta t|\lesssim 1.28$, with the maximum at 
$\eta\Delta t \approx \pm 1$. Note also 
that $|\eta\Delta t|\lesssim 0.35$ corresponds 
to an essentially conservative method, since then 
$|\mathrm{Amp}(\mathrm{i}\eta\Delta t)|$ is very 
close to unity. Therefore, assuming $\lambda$ in the model 
problem is purely imaginary, we expect growth in the 
numerical solution for timesteps $\Delta t \lesssim 
1.28|\lambda|^{-1}$, and absolute stability for 
$\Delta t \gtrsim 1.28|\lambda|^{-1}$.  
Figure~\ref{fig:ImpGRrStabilityDiagram} 
provides the analogous information for GRrSISDC2;  
the bottom plot
indicates growth for timesteps 
$\Delta t \lesssim 0.51|\lambda|^{-1}$
but absolute stability for 
$\Delta t \gtrsim 0.51|\lambda|^{-1}$. 
We now attempt to identify $\lambda$ in the model 
problem with characteristic speeds for the 
evolution system \eqref{eq:GhSystem}.

%%%%%%%%%%%%%%%%%%%%%%%%%%%%%%%%%%%%%%%%%%%%%%%%%%%%%%%%%%%%%%%%
%%%%%%%%%%%%%%%%%%%%%%%%%%%%%%%%%%%%%%%%%%%%%%%%%%%%%%%%%%%%%%%%

Given an outward-pointing unit normal $n^k$ (often to the 
boundary of a computational domain or subdomain), 
the characteristic variables of Eqs.~(\ref{eq:GhSystem}) are
\begin{equation}\label{eq:CharFields}
\psi_{ab},\quad 
\Pi_{ab} \pm n^k\Phi_{kab},\quad
(\delta^k_j - n_j n^k)\Phi_{kab},
\end{equation}
and their respective characteristic 
speeds are 
\begin{equation}\label{eq:CharSpeeds}
-n_k\shift^k,\quad -n_k\shift^k\pm N,\quad -n_k\shift^k.
\end{equation}
Equations~(\ref{eq:CharFields}) and~(\ref{eq:CharSpeeds}) 
are derived in~\cite{Lindblom2006}
[see Eqs.~(32)--(34) of that reference and the 
text thereafter, but set $\gamma_2 = 0 = \gamma_1$ as is 
the case here]. For the Schwarzschild solution in Kerr-Schild 
coordinates (see Eq.~(34) of \cite{LauPfeiffer2008}), 
the characteristic speeds for propagation orthogonal 
to an $r=\mbox{const}$ sphere reduce to
\begin{subequations}
\begin{align}\label{eq:AdvectionSpeed}
n_k V^k & = \frac{2M}{\sqrt{r^2+2Mr}},\\
n_k V^k \pm N & = \frac{2M}{\sqrt{r^2+2Mr}} \pm 
\sqrt{\frac{r}{r + 2M}},
\end{align}
\end{subequations}
where these expressions correspond to
coordinate spheres adapted to the spherical 
symmetry, i.e.~to Cartesian center $(0,0,0)$.
The smallest speeds (in magnitude) 
are $n_k V^k$ near the outer boundary ($r$ large), 
and $n_k V^k - N$ near the horizon ($r = 2M$).

%%%%%%%%%%%%%%%%%%%%%%%%%%%%%%%%%%%%%%%%%%%%%%%%%%%%%%%%%%%%%%%%
%%%%%%%%%%%%%%%%%%%%%%%%%%%%%%%%%%%%%%%%%%%%%%%%%%%%%%%%%%%%%%%%

An instability driven by the speed Eq.~(\ref{eq:AdvectionSpeed}) 
evaluated at the outer boundary appears consistent with the 
stability diagrams Figs.~\ref{fig:ImpGLoStabilityDiagram}
and~\ref{fig:ImpGRrStabilityDiagram} in the following sense: 
At the outer boundary $r=11.9$, $n_k V^k \approx 0.16$. 
Assuming wave solutions propagating with this characteristic 
speed, we have $\lambda = \mathrm{i}0.16$ in the model 
problem above.  Our simple analysis predicts instability 
when $\Delta t \lesssim 8.0$ for GLoSISDC3 and $\Delta
t \lesssim 3.2$ for GRrSISDC2, with stability for 
$\Delta t$ larger than these estimates. 
The results depicted in Figs.~\ref{fig:LongTimeErrorsGLo} and 
\ref{fig:LongTimeErrorsGRr} are consistent with these 
predictions. 

Note that the bottom panels of 
Figs.~\ref{fig:ImpGLoStabilityDiagram} 
and~\ref{fig:ImpGRrStabilityDiagram} indicate better 
stability properties for $|\eta\Delta t|$ close to 
zero.  However, even if the characteristic speeds 
at the outer boundary correspond 
to this ``near-stable'' portion of 
the imaginary axis in the relevant stability diagram, 
the characteristic speeds normal to 
$r=\mbox{const}$ surfaces for smaller radius $r$ have 
larger characteristic speeds, and  thus 
$|\mbox{Amp}(\mathrm{i}\eta\Delta t)|$ near its maximum.

Moreover, the predictions of our stability 
analysis appear at least qualitatively correct when 
the location of the outer boundary is moved to larger radii, 
where $n_kV^k$ is smaller. As $n_kV^k$ decreases, larger 
timesteps $\Delta t$ should become unstable.  Indeed, 
with GLoSISDC3 for example, we find that $\Delta t=8$ 
is unstable for $r = 18.9$ (and apparently independent of 
radial resolution). By similarly pushing the outer boundary 
outward, we can render $\Delta t = 4$ unstable for GRrSISDC2. 
Finally, we note that the standard 
stability region for backward Euler contains the entire imaginary 
axis, and is dissipative for imaginary 
$\lambda$. All of our evolutions with ImexEuler have proved 
correspondingly stable, even for small timesteps 
(with $\Delta t = 1/2$ the smallest considered).

\subsection{Convergence of the IMEX method}
We now verify both the temporal and 
spatial convergence of our scheme, using the perturbed 
initial data [case (ii)] described both above and in more
detail in Appendix~\ref{App:ID}. We continue to use
$(\gamma_0^I,\gamma_0^E) = (1,0)$, and to adopt 
exponential mappings for all radial intervals.

To verify temporal convergence, we first construct an accurate reference 
solution obtained by evolving the perturbed-black-hole initial 
data to final time $t_F = 15.0$ 
with an explicit Dormand Prince 5 (DP5) timestepper and timestep 
$\Delta t = 0.015625$. The spatial domain is determined by a top spherical 
harmonic index $\ell_\mathrm{max} = 15$ and $1.9 \leq r \leq 81.9$, and
is divided into 8 equally spaced concentric shells, each with 
with $N_r = 21$ radial collocation points.
Next, for each in a sequence of increasingly smaller timesteps we 
perform an analogous IMEX evolution using the GLoSISDC3 timestepper,
which is fourth order accurate. One complication involves boundary 
conditions: we must ensure that the choices for the explicit and IMEX 
evolutions are consistent. For both we have chosen a 
``frozen" condition, in which the incoming characteristic is fixed to 
its initial value, i.e.~we freeze $U_{ab}^-$ in 
Eq.~\eqref{eq:ImexFixUminus} to its initial value. 

\begin{figure}[t]
\includegraphics[width=0.5\textwidth]{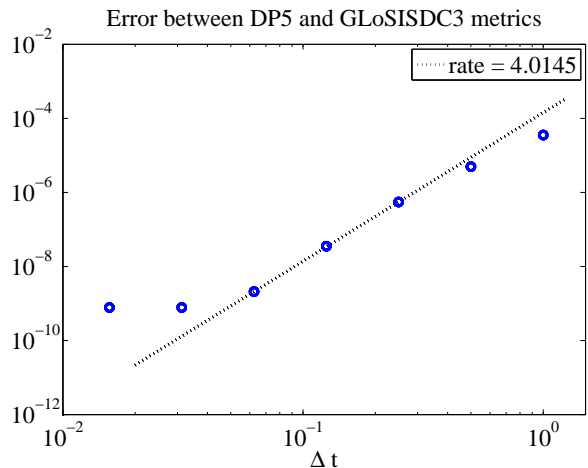}
\caption{Temporal convergence test.
Error points (circles) have been computed using \eqref{eq:psiInfErr}
in the text. The straight line in the plot and its indicated slope have 
been computed by a least squares fit of the third through fifth error 
points.
}
\label{fig:TemporalCvgTest}
\end{figure}

We compute the error,
\begin{equation}\label{eq:psiInfErr}
\|\Delta \psi\|_\infty =
\max_{a,b}\|\psi^\mathrm{GLoSISDC3}_{ab} 
          - \psi^\mathrm{DP5}_{ab}\|_\infty,
\end{equation} 
and plot it in Figure \ref{fig:TemporalCvgTest}. 
For intermediate $\Delta t$, we observe the predicted
fourth-order convergence rate. We remark that all 
timesteps shown in Fig.~\ref{fig:TemporalCvgTest}, 
except the largest, correspond to
$\Delta t \ll |\lambda|^{-1}$ from the standpoint of the
model problem analyzed in Section \ref{sec:stability}.
However, we have 
encountered no stability issues with these short-time 
evolutions.

%%%%%%%%%%%%%%%%%%%%%%%%%%%%%%%%%%%%%%%%%%%%%%%%%%%%%%%%%%%%%%%%
%%%%%%%%%%%%%%%%%%%%%%%%%%%%%%%%%%%%%%%%%%%%%%%%%%%%%%%%%%%%%%%%

%%%%%%%%%%%%%%%%%%%%%%%%%%%%%%%%%%%%%%%%%%%%%%%%%%%%%%%%%%%%%%%%
\begin{figure}
\includegraphics[width=0.4\textwidth]{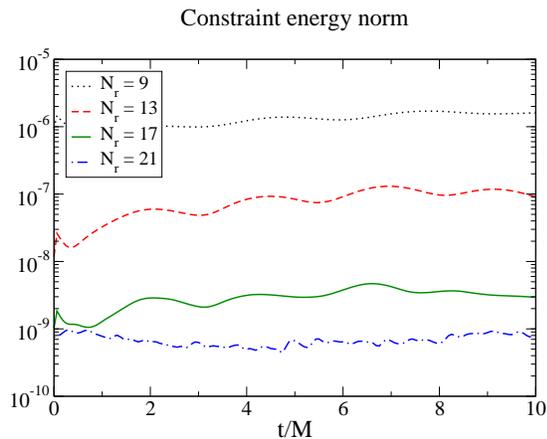}
\caption{Spatial convergence test.
This plot depicts histories for the constraint energy norm 
$\sqrt{\mathcal{E}_c}$ described in the text.
}
\label{fig:ConstraintsSpatialCvg}
\end{figure}
%%%%%%%%%%%%%%%%%%%%%%%%%%%%%%%%%%%%%%%%%%%%%%%%%%%%%%%%%%%%%%%%

We test spatial convergence as follows. Our spatial domain,
determined by $\ell_\mathrm{max} = 15$ and $1.9 \leq r \leq 41.9$,
is divided into 4 equally spaced concentric shells. For a fixed
$\Delta t = 0.0625$, we then evolve the perturbed-black-hole
initial data for different number $N_r$ of radial collocation 
points in each shell.  We compute the root-mean-square sum of 
all constraint violations $\sqrt{\mathcal{E}_c}$ (see Eq.~(53) 
of Ref.~\cite{Lindblom2006} for the precise definition), and plot 
it in Fig.~\ref{fig:ConstraintsSpatialCvg}. The figure indicates 
that the solution is dominated by spatial
error, and exhibits convergence with increased spatial resolution.
A plot of the dimensionless constraint norm $\|\mathcal{C}\|$ 
defined in Eq.~(71) of \cite{Lindblom2006} is qualitatively the
same.

%%%%%%%%%%%%%%%%%%%%%%%%%%%%%%%%%%%%%%%%%%%%%%%%%%%%%%%%%%%%%%%%
\subsection{Treatment of constraint damping terms}
%%%%%%%%%%%%%%%%%%%%%%%%%%%%%%%%%%%%%%%%%%%%%%%%%%%%%%%%%%%%%%%%

\begin{figure}
  \includegraphics[scale=0.5]{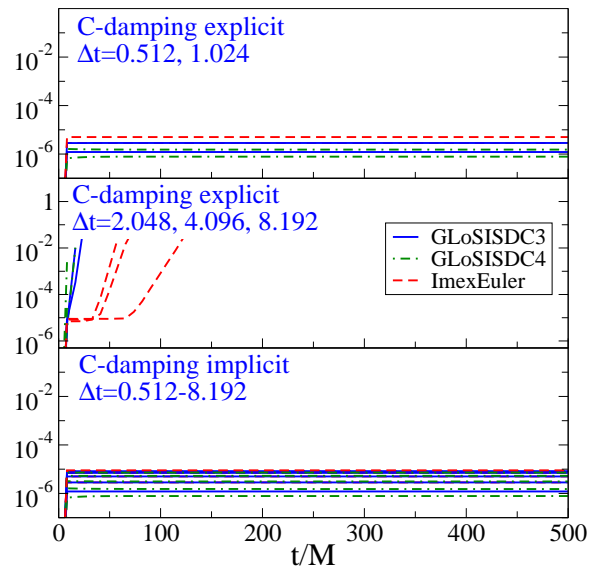}
  \caption{\label{fig:ConstraintDampingTest} Stability of various
    timesteppers when the constraint damping terms are treated
    explicitly or implicitly. Plotted are constraint violations
$\sqrt{\mathcal{E}_c}$.
The top two panels show explicit
    treatment of the constraint damping terms.  This is stable for
    small timesteps $\Delta t\le 1.024$ (top panel) and unstable for
    large timesteps, $\Delta t\ge 2.048$ (middle panel).  The lowest
    panel shows implicit treatment of the constraint damping term,
    resulting in stable evolutions for all timesteps. }
\end{figure}

As described in Sec.~\ref{sec:GH}, the generalized harmonic 
equations~(\ref{eq:GhSystem}) are modified by constraint 
damping terms proportional to $\gamma_0$ in Eq.~(\ref{eq:GH-Pi}). 
These terms cause constraint violations to decay exponentially. 
Because these terms are stiff, they require attention when choosing 
the IMEX splitting, as we now demonstrate. 

We perform runs similar to Fig.~\ref{fig:LongTimeErrorsGLo} 
but for explicit ($\gamma_0^E=1, \gamma_0^I=0$) and implicit  
($\gamma_0^E=0, \gamma_0^I=1$) constraint damping.  The 
computational domain is the same as in 
Fig.~\ref{fig:LongTimeErrorsGLo} but with Cartesian
center $(0,0,0)$, $N_r=17$, and $L=9$. Our final 
evolution time for these runs is short enough that the weak 
instabilities (associated with small GLoSISDC3 timesteps) 
observed in Fig.~\ref{fig:LongTimeErrorsGLo} do not arise.
Figure~\ref{fig:ConstraintDampingTest} shows the constraints 
for various timesteps and three different IMEX timesteppers.  
From the lowest panel, we see that the system is well-behaved 
for all considered timesteps if the constraint-damping terms are 
treated {\em implicitly}.  The upper two panels show that for 
{\em explicit} handling of the constraint damping terms, the 
timestep matters: For small $\Delta t$, the simulations
behave well, for large $\Delta t$ they blow up.  This is 
consistent with a Courant limit for the explicit sector of 
the timestepper, arising from the constraint-damping term.

%%%%%%%%%%%%%%%%%%%%%%%%%%%%%%%%%%%%%%%%%%%%%%%%%%%%%%%%%%%%%%%%
\subsection{Adaptive timestepping and comparison to explicit timestepper}
%%%%%%%%%%%%%%%%%%%%%%%%%%%%%%%%%%%%%%%%%%%%%%%%%%%%%%%%%%%%%%%%

%%%%%%%%%%%%%%%%%%%%%%%%%%%%%%%%%%%%%%%%%%%%%%%%%%%%%%%%%%%%%%%%
\begin{figure}
\includegraphics[width=0.45\textwidth]{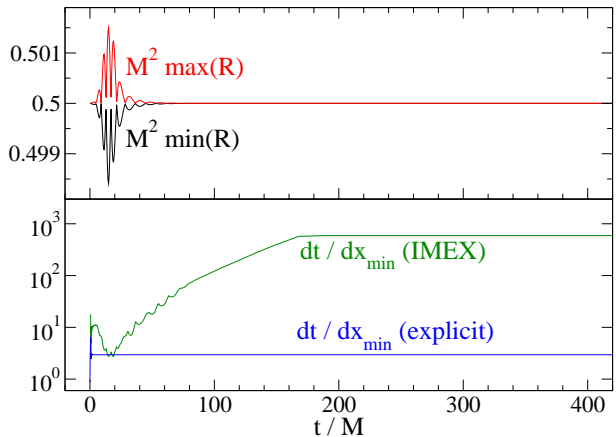}
\caption{Demonstration of IMEX evolution of a single perturbed black hole 
using GRrSISDC2 with 
adaptive timestepping. 
\emph{Top panel:} the minimum and maximum of the horizon's 
dimensionless intrinsic scalar curvature $M^2 R$, which characterizes 
the horizon shape. 
\emph{Bottom panel:} The Courant factor $\Delta t / \Delta x_{\rm min}$, 
where $\Delta t$ is the 
size of each timestep and $\Delta x_{\rm min}$ is the minimum spacing 
between grid-points, for an IMEX evolution and for an analogous explicit 
evolution.
Both evolutions are evolved at the same spatial resolution (with approximately 
$43^3$ grid-points).
\label{fig:Adapt_ShapeAnddt}}
\end{figure}
%%%%%%%%%%%%%%%%%%%%%%%%%%%%%%%%%%%%%%%%%%%%%%%%%%%%%%%%%%%%%%%%

In this subsection, we demonstrate adaptive timestepping in an IMEX 
evolution by using an adaptive timestepper on the perturbed-black-hole 
initial data from Appendix~\ref{App:ID}.
We evolved this initial data on a set of 16 concentric spherical shells 
with Cartesian center (0,0,0) and with $1.9 \leq r \leq 161.9$, 
$N_r=17$, and $L=11$. A gravitational-wave pulse falls into a 
nonspinning black hole of mass $M=1$ shortly after $t=0$, which 
causes a time-dependent deformation of the hole's horizon. 
The top panel of Fig.~\ref{fig:Adapt_ShapeAnddt} shows the minimum 
and maximum values of the intrinsic scalar curvature $R$
of the horizon: As the wave falls into the hole, the horizon shape 
oscillates and then relaxes back to the Schwarzschild value 
$M^2 R = 1/2$, which holds for the 
curvature of a sphere of Schwarzschild radius $r=2M$.

The bottom panel of Fig.~\ref{fig:Adapt_ShapeAnddt} 
plots the step size chosen by the adaptive timestepper 
$\Delta t/\Delta x_{\rm min}$ for an IMEX evolution and an analogous explicit evolution 
of the same initial data. The explicit timestepper chooses an essentially 
constant $\Delta t$, right at its CFL stability limit.
During the initial perturbation, the IMEX step size decreases 
to a local minimum; as the hole relaxes to its 
final time-independent configuration, the step size increases, eventually 
reaching an artificially imposed upper limit. (This upper limit was chosen 
to guarantee that the elliptic solver would converge in a reasonable amount 
of wallclock time.)

During the initial time-dependent perturbation, the IMEX evolution 
is usually able to take significantly larger timesteps than the 
analogous explicit 
evolution. 
In the explicit evolution, the 
Courant factor is limited to $\Delta t/\Delta x_{\rm min}\approx 3$, which is 
comparable to the minimum of the IMEX evolution's Courant factor.

We remark that the above IMEX simulations exhibit 
some instability: the IMEX run shows slow constraint growth,
perhaps 
because we did not impose a constraint-preserving boundary condition on the 
outer boundary.
However, the analogous explicit evolution exhibits no instability, 
and the IMEX 
and explicit evolutions' constraint violations are comparable in size 
when we terminate the simulations (after time $t=2000 M$, 
which is long after the spacetime has 
relaxed to its final, stationary state).

%%%%%%%%%%%%%%%%%%%%%%%%%%%%%%%%%%%%%%%%%%%%%%%%%%%%%%%%%%%%%%%%
\section{Discussion}
\label{sec:Discussion}
%%%%%%%%%%%%%%%%%%%%%%%%%%%%%%%%%%%%%%%%%%%%%%%%%%%%%%%%%%%%%%%%

\subsection{Results obtained in the present work}
\label{sec:DiscussionA}

In this article, we have further developed IMEX-techniques applied to
hyperbolic systems.  Specifically, we have moved beyond the model
problem of a scalar wave~\cite{LauPfeiffer2008} to the study of the
full non-linear Einstein's equations for single black hole spacetimes.
Many results of the model problem presented in~\cite{LauPfeiffer2008}
carry over to Einstein's equations in generalized 
harmonic form~\cite{Lindblom2006}:
We continue to rewrite the implicit equation in second order form to
utilize an existing elliptic solver~\cite{Pfeiffer2003}.  Furthermore,
as in the scalar-field case, we do not impose a boundary condition at
the excision boundary inside the black hole.  Uniqueness of the solution
of the second order implicit equation is enforced, we believe, 
by the demand that the solution be {\em regular} across the horizon.

In contrast to the model problem, the generalized harmonic evolution
system contains physical constraints\footnote{These are in addition to
the auxiliary constraints arising from the reduction to 
first order form.} which in explicit simulations are handled with 
constraint damping~\cite{Gundlach2005,Pretorius2005a,Lindblom2006}.  
We have introduced
analogous constraint damping terms in the IMEX formulation, namely the
terms proportional to $\gamma_0^I$ in 
Eqs.~(\ref{eq:ImexGhCase2})
and~(\ref{eq:QandG}).  We have found that these constraint damping terms are
essential for stability.  Treating the constraint damping terms
explicitly incurs a Courant limit due to their stiffness, and so we
recommend an implicit treatment of these terms ($\gamma_0^E=0;
\gamma_0^I=\gamma_0$).

We have focused our investigation on spectral deferred correction
schemes~\cite{Dutt2000,Minion2003,LaytonMinion,HagstromZhou2006}, 
utilizing 3 Gauss-Lobatto and 2 Gauss-Radau-right quadrature points:
GLoSISDC3 and GRrSISDC2, respectively.  These schemes
generally work well; however, we find a weak
instability for small timesteps which may be 
related to the stability region of the implicit sector of these 
IMEX schemes.  We also have investigated
ImexEuler and third order Additive 
Runge Kutta (ARK3). While ImexEuler 
proved robustly stable, our simulations with ARK3 showed a linear 
growing instability. The origin of this instability remains an 
open question.

The most demanding scenario that we have considered is a perturbed 
single black hole that rings down to a quiescent state.  We have 
evolved this configuration with explicit and IMEX techniques. The 
explicit evolution used a fifth order Dormand-Prince timestepper 
with adaptive timestepping; however, because of the 
necessarily small grid-spacing 
close to the black hole, the explicit simulation uses an essentially
constant timestep at its Courant limit, cf.~Fig.~\ref{fig:Adapt_ShapeAnddt}. 
The IMEX method uses a small timestep for the early, dynamic part of 
the simulation, and then chooses increasingly larger timesteps, until 
it exceeds the explicit timestep by about a factor of 200.  

For very large timesteps, the convergence rate of our elliptic solver
deteriorates, and overall efficiency drops.  Therefore, so far we have
limited the IMEX timestep to $\approx 200$ times the explicit timestep.
For these timesteps, the computational efficiency of the implicit and
explicit code are approximately similar, for the example shown in
Fig.~\ref{fig:Adapt_ShapeAnddt}.  We are confident that improved
preconditioning will accelerate convergence of the implicit solver,
allowing us to utilize yet larger timesteps in IMEX at lower
computational cost.  Besides improved preconditioning,  
several aspects of our future work will increase the efficiency
of the IMEX code: We plan to implement a more accurate 
starting method for 
the prediction phase of an SISDC timestep.  
We further plan to perform a detailed analysis of the required tolerances 
in the implicit solve (in the present work we set tolerances near
numerical round-off to eliminate spurious instabilities due to
insufficient accuracy), and we plan to optimize the C++ code 
implementing Eq.~(\ref{eq:psiEquation}).  We expect these steps to 
significantly increase efficiency of the IMEX code;  in contrast, 
the explicit code is already highly optimized.  
In the next subsection, we discuss additional code improvements 
relevant to IMEX evolutions of binary black holes.

%%%%%%%%%%%%%%%%%%%%%%%%%%%%%%%%%%%%%%%%%%%%%%%%%%%%%%%%%%%%%%%%
\subsection{Prospects for binary black hole evolutions}
%%%%%%%%%%%%%%%%%%%%%%%%%%%%%%%%%%%%%%%%%%%%%%%%%%%%%%%%%%%%%%%%
\label{sec:4B}

Long and accurate binary black hole simulations are needed for optimal
signal-processing of current and future gravitational
wave-detectors~\cite{Hannam:2010,Damour:2010,MacDonald:2011ne,Boyle:2011dy};
this provides the motivation for the present work.  While the
results obtained here are very encouraging, additional work will be
necessary to apply IMEX to black hole binaries.

First, the formalism must be adopted to the dual-frame
  approach~\cite{Scheel2006} used in binary black hole simulations
  with {\tt SpEC}.  The corotating coordinates implemented via the
  dual-frame technique are essential for implicit time-stepping,
  because they localize the black holes in the computational
  coordinates.  Without corotating coordinates, the black holes would
  move across the grid, resulting in rapid time-variability of the
  solution (on timescales $M/v$, where $v$ denotes the velocity of
  the black hole with mass $M$).  This variability would necessitate a
  small time-step to achieve small time-discretization error. 
The dual-frame technique merely adds a new advection term
  into the evolution equations, therefore, we expect the extension to
  dual-frames to be straightforward. 

Second, the implicit solver must remain efficient despite the
  more complicated computational domain.  And third, good outer
  boundary conditions will be necessary.  We expect that 
  the second and third
  issues can be addressed simultaneously with the following ideas:
{\tt SpEC} evolves binary black holes on a domain decomposition
consisting of ``inner'' spherical shells around each of the black
holes, which are surrounded by a complicated structure of ``outer''
subdomains (cylinders, distorted blocks and spherical shells, the
latter of which extend to a large outer radius).  The inner spherical
shells require the highest resolution and therefore determine the
Courant condition for fully explicit evolutions.

To simulate binary black holes
with IMEX methods, we envision a split-by-region approach
\cite{Kanevsky2007}, where the inner spherical shells are treated
with the IMEX techniques described in this paper and the outer
subdomains are handled explicitly.  The split-by-region approach has
two important advantages: First, implicit equations will have to be
solved only on series of concentric shells.  This is the case
considered here, for which {\tt SpEC}'s elliptic solver is already
reasonably efficient with further possible efficiency improvements as
discussed in Sec.~\ref{sec:DiscussionA}.  In contrast, solution of
implicit equations on the entire (rather complicated)
domain-decomposition would likely be less efficient because of
difficulties in preconditioning the inter-subdomain boundary
conditions.  Second, for explicit evolutions non-reflecting and
constraint-preserving outer boundary conditions are
available~\cite{Lindblom2006,Rinne2007,Rinne2008b}.  Explicit
treatment of the region near the outer boundary will allow us to reuse
these boundary conditions.  In contrast, similarly sophisticated
boundary conditions have not yet been investigated in an IMEX setting.

Because the outer subdomains will be handled
explicitly, the split-by-region scheme will still be subject to a
Courant condition, based on the minimum grid-spacing $\Delta x_{\rm outer}$ in
the explicitly evolved region.  Because the minimum grid-spacing in the outer subdomains is larger than the minimum grid-spacing $\Delta x_{\rm inner}$ near the
black holes, the envisioned split-by-region approach should
allow for timesteps larger by a factor
\begin{equation}\label{eq:RDeltat}
R_{\rm \Delta t}
\equiv
\frac{\Delta x_{\rm outer}}{\Delta x_{\rm inner}}\gg 1.
\end{equation}
We shall assume that the cost-per-timestep is proportional to 
the number of collocation points, with different constants 
for explicit
and IMEX cases:
\begin{align}
  C_{\rm explicit} &= C(N_{\rm outer}+N_{\rm inner})\\
  C_{\rm IMEX} &= C N_{\rm outer} + C R_{\rm step} N_{\rm inner}
\end{align}
Here, $R_{\rm step}$ is the ratio of the cost of an IMEX-timestep to a
fully explicit timestep.  The simulations presented in
Sec.~\ref{sec:NumericalExperiments} give $R_{\rm step} \approx 
100$, with $R_{\rm step}$ being somewhat larger for very large 
$\Delta t$ and somewhat smaller for small $\Delta t$.

For temporal integration to a fixed final time, the number of
timesteps for a fully explicit scheme will be proportional to
$1/\Delta x_{\rm inner}$, whereas for the IMEX split-by-region
scheme, the number of timesteps will be proportional to $1/\Delta
x_{\rm outer}$.  Therefore, the IMEX split-by-region scheme should
require the following fractional amount of CPU resources relative to a
completely explicit evolution (a smaller number indicates advantage
for IMEX):
\begin{align}
R_{\rm BBH} 
\equiv 
\frac{\Delta x_{\rm inner}}{\Delta x_{\rm outer}}
\frac{C_{\rm IMEX} }{C_{\rm explicit}} 
=\frac{1}{R_{\rm \Delta t}}
\frac{N_{\rm outer}+R_{\rm step}N_{\rm inner}}{N_{\rm outer}+N_{\rm inner}}.
\end{align}
When $R_{\rm step}N_{\rm inner} \gg N_{\rm outer}$, this simplifies to
\begin{equation}
R_{\rm BBH} 
\approx\frac{R_{\rm step}}{R_{\Delta t}}\,\frac{N_{\rm inner}}{N_{\rm inner}+N_{\rm outer}}.
\label{eq:BBH-cost}
\end{equation}
As expected, the question is whether the larger 
timestep, encoded in $R_{\rm \Delta t}$, can 
compensate for the additional cost per timestep, 
encoded in $R_{\rm step}$.  However, split-by-region 
mitigates the effect of $R_{\rm step}$ by 
an extra factor $N_{\rm inner}/N_{\rm total}$.

To make this discussion concrete, a recent mass-ratio 
$q\!=\!6$ simulation of non-spinning black holes used
$N_{\rm outer}\!=\!219222$, $N_{\rm inner}\!=\!147288$, 
and $R_{\Delta t}\!=\!34$. With these values 
Eq.~(\ref{eq:BBH-cost}) gives $R_{\rm BBH} = 1.2.$, 
i.e.~an IMEX evolution should be marginally 
more expensive than a fully explicit one.
As the mass-ratio is further increased,
the grid-spacing needed to resolve the smaller black 
hole decreases proportionally.  Therefore,
$\Delta x_{\rm inner}$ will decrease proportional 
to $1/q$, and $R_{\Delta t}$ will increase proportional 
to $q$.  The constant of proportionality can be determined 
from $R_{\Delta t}=34$
at $q=6$, so that $R_{\Delta t} \approx 6q$. 
The number of grid-points will only 
modestly change, so we assume $N_{\rm inner} \approx 
N_{\rm outer}$. Then from Eq.~(\ref{eq:BBH-cost}) we 
estimate an efficiency increase for IMEX of
\begin{equation}
R_{\rm BBH} \approx \frac{100}{6q}\,\frac{1}{2}\approx \frac{8}{q}.
\end{equation}
Therefore, with increasing mass-ratio, IMEX will become 
increasingly more efficient than the explicit 
evolution code.

The additional efficiency gains for IMEX discussed in
Sec.~\ref{sec:DiscussionA} are not taken into account in this
estimate.  Furthermore, a more judicious choice of domain
decomposition with a more carefully tuned number of collocation points
in the inner spheres would reduce the ratio $N_{\rm inner}/N_{\rm
  total}$.  Finally, we have not accounted for the fact that BBH
evolutions require additional CPU resources for interpolation.
Because interpolation occurs only in the outer subdomains, this will
reduce $R_{\rm step}$.

On the other hand, at this point we do not know how accurately the
implicit equations must be solved in the binary case; if higher
accuracy is required to control secularly accumulating phase-errors,
then each implicit solve would become more expensive.  Furthermore,
the binary simulations utilize a dual-frame method which will add some
overhead to the implicit solutions.  

In summary, we believe that IMEX schemes offer the
promise of faster binary black-hole simulations, but many 
interesting issues (such as those outlined in this section)
deserve further investigation.

%%%%%%%%%%%%%%%%%%%%%%%%%%%%%%%%%%%%%%%%%%%%%%%%%%%%%%%%%%%%%%%%

\subsection{Applicability to other computational techniques}

  The results in this paper were obtained for the generalized
  harmonic formulation of Einstein's equations using pseudo-spectral
  methods.  IMEX methods might also be implemented for other
  formulations of the Einstein equations, such as the 
  Baumgarte-Shapiro-Shibata-Nakamura
  (BSSN) formulation
  \cite{shibata95,Baumgarte1998} or the recent conformal
  decompositions of the Z4 formulation \cite{Alic:2011gg}. Indeed, for 
  such systems specification of the first-order implicit system 
  [analogous to Eqs.~\eqref{eq:ImexGhCase2}] corresponding to a single 
  time-step is straightforward. However, relative to the analogous 
  reduction performed for the generalized harmonic formulation in this 
  paper, the reduction of such a first-order system to a second-order 
  system involving, presumably, some subset of the system variables 
  would seem to be more involved.  A second 
  impediment arises from the need to use corotating coordinates. In 
  corotating coordinates, temporal timescales are long, allowing large 
  time-steps with sufficiently small time-discretization error
  (cf. Sec.~\ref{sec:4B}).  To our knowledge, none of the BSSN/Z4
  codes currently utilize corotating coordinates, although, in
  principle, the dual-frame approach~\cite{Scheel2006} could be 
  applied in such codes. 

Provided the existence of efficient
solvers for the resulting discretized implicit equations, the IMEX
methods developed here should also be applicable to other spatial
discretizations, e.g.~finite differences, finite elements, or other
Galerkin spectral-element approaches. 
The presence
of a horizon and the replacement of an inner boundary condition by a
regularity condition (cf. Sec.~\ref{sec:BC}) are points demanding
particular attention.  In our approach each component of the apparent
horizon is covered by a single subdomain. Therefore, in our
pseudo-spectral treatment the metric in the vicinity of the horizon is
expanded in terms of a single set of basis functions, with regularity
of the solution an automatic consequence. Guaranteed regularity of the
solution might be lost for either a finite-difference method or an
unstructured-mesh method, but further studies of these possibilities
are clearly warranted.

%%%%%%%%%%%%%%%%%%%%%%%%%%%%%%%%%%%%%%%%%%%%%%%%%%%%%%%%%%%%%%%%

\acknowledgments 

We are pleased to thank Saul Teukolsky, Larry Kidder, 
Jan Hesthaven, and Mike Minion for helpful discussions. 
This work was supported in part by the Sherman Fairchild foundation, 
NSF grants Nos.~PHY-0969111 and PHY-1005426, and NASA grant 
No.~NNX09AF96G at Cornell; 
and by NSF grant No.~PHY 0855678 to the University of New Mexico.
H.P. gratefully acknowledges support from the NSERC of Canada, from 
the Canada Research Chairs Program, and from the Canadian Institute 
for Advanced Research. Some computations in this paper were performed using 
the GPC supercomputer at the SciNet HPC Consortium; SciNet is funded by: the 
Canada Foundation for Innovation under the auspices of Compute Canada; the 
Government of Ontario; Ontario Research Fund --- Research Excellence; and 
the University of Toronto. Some computations in this paper were performed 
using Pequena at the UNM Center for Advanced Research Computing.

\appendix
\section{Decomposition of $\Gamma_e$}\label{sec:GammaDecomp}
The trace $\psi^{ab}\Gamma_{eab}$ of the Christoffel symbol 
$\Gamma_{eab}$ of the first kind is
\begin{equation}
\Gamma_e = \psi^{ab} \partial_a \psi_{eb} 
- \frac{1}{2}\psi^{ab} \partial_e \psi_{ab}.
\end{equation}
Writing the time-derivative separately, we reach
\begin{align}
\Gamma_e  =& 
\psi^{0b} \partial_0  \psi_{eb}
+ \psi^{kb} \partial_k \psi_{eb}
\nonumber \\
& - \frac{1}{2}\psi^{ab} \delta^0_e \partial_0 \psi_{ab}
  - \frac{1}{2}\psi^{ab} \delta^k_e\partial_k \psi_{ab},
\end{align}
where $0$ is the time $t$ component. Now we insert the 
identities $\partial_t \psi_{ab}
= -N\Pi_{ab} + V^k \Phi_{k ab}$ and
$\partial_k \psi_{ab} = \Phi_{k ab}$, thereby
finding
\begin{align}
\Gamma_e & =
-N\psi^{0b} \Pi_{eb}
+ \psi^{0b} V^k \Phi_{k eb}
+ \psi^{k b} \Phi_{k eb}
\nonumber \\
& + \frac{1}{2}N\psi^{ab} \psi^0_e \Pi_{ab}
  - \frac{1}{2}\psi^{ab} \psi^0_e V^k \Phi_{k ab}
  - \frac{1}{2}\psi^{ab} \psi^k_e \Phi_{k ab}.
\end{align}
Finally, we use the identity $t^a = -N\psi^{0a}$ to write
\begin{align}
\Gamma_e & =
t^b \Pi_{eb}
- N^{-1} V^k t^b \Phi_{k eb}
+ \psi^{k b} \Phi_{k eb}
\nonumber \\
& - \frac{1}{2}t_e \psi^{ab} \Pi_{ab}
  + \frac{1}{2}N^{-1}V^k \psi^{ab} t_e \Phi_{k ab}
  - \frac{1}{2}\psi^{ab} \psi^k_e \Phi_{k ab}.
\end{align}
Using the last expression, we compute
\begin{subequations}
\begin{align}
\frac{\partial\Gamma_e}{\partial\Pi_{cd}}
& = \frac{1}{2}t^c \psi^d_e
  + \frac{1}{2}t^d \psi^c_e
  - \frac{1}{2}t_e \psi^{cd}
\\
\Gamma_e - \frac{\partial\Gamma_e}{\partial\Pi_{cd}}\Pi_{cd}
 & = \big(\psi^{k b}
- N^{-1} V^k t^b\big)\Phi_{k eb}
\nonumber \\
& - \frac{1}{2}\big(\psi^k_e - N^{-1}V^k t_e\big)
\psi^{ab} \Phi_{k ab},
\end{align}
\end{subequations}
and these formulas complete the definitions in Eqs.~(\ref{eq:QandG}).
%Then
%\begin{subequations}
%\begin{align}
%\mathcal{Q}_{ab}{}^{cd}
%& = \gamma_0\big[
% 2t^{(c} t_{(a} \psi^{d)}_{b)}
%- t_a t_b \psi^{cd} 
%- \psi_{ab} t^c t^d
%- \frac{1}{2}\psi_{ab}\psi^{cd}\big]\\
%\mathcal{G}_{ab}
%& = \gamma_0 (\psi^{kc} - N^{-1} V^k t^c)(2t_{(a}\Phi_{|k|b)c}
%- \psi_{ab}t^d \Phi_{kcd})
%\nonumber \\
%& - \frac{1}{2}\gamma_0(2 t_{(a}\psi^k_{b)}
%- 2 t_a t_b N^{-1} V^k)\psi^{cd}\Phi_{kcd}
%\end{align}
%\end{subequations}

\section{Semi-implicit spectral deferred corrections}\label{sec:SISDC}
This appendix describes one of the IMEX timestepping 
algorithm used for our evolutions, summarizing results found 
in Refs.~\cite{Dutt2000,Minion2003,LaytonMinion,HagstromZhou2006} 
and expressing them in our notation. We aim here only to
describe the algorithm, and do not address stability and
convergence issues (which have been exhaustively explored 
in the references).

\subsection{Collocation approximation of the Picard 
integral} We start with the generic ODE initial 
value problem Eq.~(\ref{eq:IVP}). Each spectral deferred 
correction method specifies a rule for advancing 
the vector $\boldsymbol{u}_n$ at the present timestep 
$t_n$ (perhaps the initial time $t_0$) to a vector 
$\boldsymbol{u}_{n+1}$ at the next timestep $t_{n+1}
= t_n + \Delta t$. The Picard integral form of the 
initial value problem Eq.~(\ref{eq:IVP}) for starting 
value $\boldsymbol{u}_n$ is
\begin{equation}\label{eq:Picard}
\boldsymbol{u}(t) =  \boldsymbol{u}_n
+ \int_{t_n}^t 
\boldsymbol{f}(s,\boldsymbol{u}(s)) ds.
\end{equation}
We consider this equation on the interval $[t_n,t_{n+1}]$, 
and show how iterative approximation of (\ref{eq:Picard}) 
yields a timestepping scheme.

Introduce $p$ collocation nodes which are also 
time sub-steps:
\begin{equation}
t_{(m)} = t_n + c_m \Delta t,
\qquad 
0 \leq c_1 < c_2 < \cdots < c_p \leq 1.
\end{equation}
The $c_m$ are either Gauss-Legendre, Gauss-Lobatto, 
or Gauss-Radau nodes relative to the standard interval 
$[0,1]$. Each of the endpoints, $t_n$ and $t_{n+1}$, 
may or may not be a collocation node. In particular, 
for the Gauss-Legendre case both $t_n$ and $t_{n+1}$ 
are not in $\{t_{(1)},\dots,t_{(p)}\}$.

Define a system vector $\boldsymbol{u}_{(m)}$ at 
each collocation point $t_{(m)}$. A solution to the
polynomial collocation approximation to the Picard 
integral (\ref{eq:Picard}) is a set 
$\{\boldsymbol{u}_{(m)} : m = 1,\dots,p\}$ of 
vectors obeying
\begin{align}\label{eq:collocPicard}
\begin{split} 
\boldsymbol{u}_{(m)} & = \boldsymbol{u}_n
+ \int_{t_n}^{t_{(m)}} \boldsymbol{\psi}(t) dt
\\
                   & = \boldsymbol{u}_n + 
\Delta t \sum_{q = 1}^p 
S_{mq}\boldsymbol{f}(t_{(q)},\boldsymbol{u}_{(q)}),\quad
m = 1,\dots, p,
\end{split}
\end{align}
where $\boldsymbol{\psi}(t)$ is the unique degree $p-1$
(vector-valued) polynomial which interpolates the data
\begin{align}
\big(t_{(m)}, 
\boldsymbol{f}(t_{(m)},\boldsymbol{u}_{(m)})\big),
\quad m = 1,\ldots, p.
\end{align}
The elements $S_{mq}$ define the spectral integration matrix. 
The solution $\{\boldsymbol{u}_{(m)} : m = 1,\dots,p\}$
to Eq.~(\ref{eq:collocPicard}) defines the approximation
\begin{equation}
\boldsymbol{u}_n
+ \int_{t_n}^{t_{n+1}} \boldsymbol{\psi}(t) dt
\approx \boldsymbol{u}(t_{n+1}).
\end{equation}
We get an {\em approximation} to the solution  
of the collocation equations (\ref{eq:collocPicard}) 
via an iteration described below (that is, we get
an approximate solution to the approximating system
of equations). 

\subsection{Iterative solution of the collocation equations}
Our iterative scheme for solving (\ref{eq:collocPicard}) relies 
on two phases: (i) an initial {\em prediction phase} which 
generates a provisional solution $\boldsymbol{u}_{(m)}^{0}$, 
and (ii) a {\em correction phase} which generates successive 
improvements $\boldsymbol{u}_{(m)}^{k}$, $k = 1,\ldots,K$ 
as described in detail below. As described by Hagstrom and
Zhou \cite{HagstromZhou2006}, the prediction phase requires a {\em
starting method}, and we use ImexEuler. For each $k$ the set 
$\{\boldsymbol{u}_{(m)}^{k} : m = 1,\dots,p\}$ determines
an interpolating polynomial $\boldsymbol{\psi}^{k}(t)$, and 
the numerically computed approximation to $\boldsymbol{u}(t_{n+1})$ is
\begin{equation}\label{eq:SISDCKun}
\boldsymbol{u}_{n+1} = \boldsymbol{u}_n
+ \int_{t_n}^{t_{n+1}} \boldsymbol{\psi}^{K}(t) dt.
\end{equation}
For the Gauss-Legendre,	Gauss-Radau-right, and Gauss-Lobatto 
cases, Hagstrom and Zhou \cite{HagstromZhou2006} have
studied the accuracy of these methods. When considered as
global methods (integration to a fixed time with 
multiple timesteps), they have shown that for sufficiently 
large $K$ the optimal order of attainable accuracy is 
respectively $2p$, $2p-1$, and $2p-2$, that is the same 
order as for the underlying quadrature rule; however, 
this order is typically not obtained for the vectors 
$\boldsymbol{u}_{(m)}^{K}$ at intermediate times. 

Typically $K = 2p-1$ for Gauss-Legendre, 
$K = 2p-2$ for Gauss-Radau (left or right), and 
$K=2p-3$ for Gauss-Lobatto cases, where each
choice should yield the optimal order of accuracy.
Our presentation of the iteration algorithm makes 
use of the notations
\begin{align}
\begin{split}
\boldsymbol{f}_{(m)}^{k}
& = \boldsymbol{f}(t_{(m)},\boldsymbol{u}_{(m)}^{k})
\\
\Delta t_{0} & = c_1\Delta t\\
\Delta t_{m} & = (c_{m+1}-c_m)\Delta t,\quad m = 
1,\dots,p-1,\\
\end{split}
\end{align}
but draws a distinction between two cases
(i) Gauss-Legendre and Gauss-Radau-right (for these
methods $t_n$ is not a collocation point) and (ii) 
Gauss-Lobatto and Gauss-Radau-left (for these
$t_n$ is a collocation point).

To start the prediction phase for Gauss-Legendre 
and Gauss-Radau-right, we first solve
\begin{align}
\boldsymbol{u}_{(1)}^{0} - \Delta t_0 
\boldsymbol{f}^I(t_{(1)},\boldsymbol{u}_{(1)}^{0}) 
= 
\boldsymbol{u}_n + \Delta t_0 
\boldsymbol{f}^E(t_n,\boldsymbol{u}_n)
\end{align}
to get $\boldsymbol{u}_{(1)}^0$. For Gauss-Lobatto or 
Gauss-Radau-left, we have $\boldsymbol{u}_{(1)}^0
=\boldsymbol{u}(t_n)$ to start with.
We then march forward in time by solving in sequence 
the following equations:
\begin{align}
\boldsymbol{u}_{(m+1)}^{0} & - 
\Delta t_{m}
\boldsymbol{f}^I(t_{(m+1)},\boldsymbol{u}_{(m+1)}^{0}) 
\nonumber \\
& = \boldsymbol{u}_{(m)}^{0} + \Delta t_{m} 
\boldsymbol{f}^E(t_{(m)},\boldsymbol{u}_{(m)}^{0})
\end{align}
for $m = 1,\dots,p-1$.
Note that each such equation is defined by the 
previously constructed $\boldsymbol{u}_{(m)}^{0}$ 
and amounts to an ImexEuler timestep.

We have used ImexEuler to generate the provisional
solution $\{\boldsymbol{u}_{(m)}^0: m = 1,\dots,p\}$,
and this simple method is also the basis of the 
correction phase. Given $\{\boldsymbol{u}_{(m)}^{k}: 
m = 1,\dots,p\}$, a correction sweep yields updated 
vectors. $\{\boldsymbol{u}_{(m)}^{k+1} : m=1,\dots,p\}$. 
To understand the eventual scheme which produces 
the updated vectors, first consider an approximate 
solution $\boldsymbol{v}(t)$ to the continuum 
initial value problem Eq.~(\ref{eq:IVP}), 
assuming $\boldsymbol{v}(t_n) = \boldsymbol{u}(t_n) = 
\boldsymbol{u}_n$, along with the residual
\begin{equation}
\boldsymbol{r}(t) = \boldsymbol{u}_n +
\int_{t_n}^t f(s,\boldsymbol{v}(s))ds - \boldsymbol{v}(t).
\end{equation}
If the exact solution is $\boldsymbol{u}(t) = 
\boldsymbol{v}(t) + \boldsymbol{\delta}(t)$, then
the correction $\boldsymbol{\delta}(t)$ obeys
\begin{equation}
\frac{d \boldsymbol{\delta}}{dt} =
\boldsymbol{f}(t,\boldsymbol{v}+\boldsymbol{\delta}) 
- \boldsymbol{f}(t,\boldsymbol{v})
+ \frac{d \boldsymbol{r}}{dt},\qquad
\boldsymbol{\delta}(t_n) = 0.
\end{equation}

We timestep this equation using ImexEuler. For
case (i), either Gauss-Legendre or Gauss-Radau-right, 
we first solve 
\begin{align}\label{eq:stepdelta1}
\begin{split}
\boldsymbol{\delta}_{(1)}
- \Delta t_0 \boldsymbol{f}^I(t_{(1)},\boldsymbol{v}_{(1)}+
\boldsymbol{\delta}_{(1)})
= - \Delta t_0 \boldsymbol{f}^I(t_{(1)},\boldsymbol{v}_{(1)})
+ \boldsymbol{r}_{(1)}
\end{split}
\end{align}
for $\boldsymbol{\delta}_{(1)}$, where to reach this
equation we have used $\boldsymbol{\delta}_{(0)} = 0 
= \boldsymbol{r}_{(0)}$. For the case (ii) methods we 
have $\boldsymbol{\delta}_{(1)} = 0$ to start with.
Subsequently, we solve
\begin{align}\label{eq:stepdelta}
\begin{split}
& \boldsymbol{\delta}_{(m+1)}
- \Delta t_m \boldsymbol{f}^I(t_{(m+1)},
\boldsymbol{v}_{(m+1)}+\boldsymbol{\delta}_{(m+1)}) \\
& = \boldsymbol{\delta}_{(m)} + 
\Delta t_m \big[\boldsymbol{f}^E(t_{(m)},\boldsymbol{v}_{(m)}+
\boldsymbol{\delta}_{(m)})
- \boldsymbol{f}^E(t_{(m)},\boldsymbol{v}_{(m)})\big]
\\
& - \Delta t_m \boldsymbol{f}^I(t_{(m+1)},\boldsymbol{v}_{(m+1)})
+ \boldsymbol{r}_{(m+1)}
- \boldsymbol{r}_{(m)}
\end{split}
\end{align}
for $m=1,\dots,p-1$.

To exploit formulas (\ref{eq:stepdelta1})--(\ref{eq:stepdelta}),
first make the assignments
\begin{align}\label{eq:assignments}
\begin{split}
\boldsymbol{\delta}_{(m)} & \rightarrow \boldsymbol{\delta}^k_{(m)}\\
\boldsymbol{v}_{(m)} & \rightarrow \boldsymbol{u}_{(m)}^k\\
\boldsymbol{v}_{(m)}+\boldsymbol{\delta}_{(m)} &
\rightarrow \boldsymbol{u}_{(m)}^{k+1} = 
\boldsymbol{u}_{(m)}^k + \boldsymbol{\delta}^k_{(m)}\\
\boldsymbol{r}_{(m)} & \rightarrow 
\boldsymbol{r}^k_{(m)} = \boldsymbol{u}_n +
\Delta t \sum_{q = 1}^p
S_{mq}\boldsymbol{f}_{(q)}^k - \boldsymbol{u}^k_{(m)}.
\end{split}
\end{align}
With these assignments in Eq.~(\ref{eq:stepdelta1}), we find,
upon adding $\boldsymbol{u}^k_{(1)}$ to both sides of
the equation,
\begin{align}
\begin{split}
\boldsymbol{u}_{(1)}^{k+1} & -
\Delta t_0
\boldsymbol{f}^I(t_{(1)},\boldsymbol{u}_{(1)}^{k+1}) 
\\
& = \boldsymbol{u}_n
- \Delta t_0 
\boldsymbol{f}^I(t_{(1)},\boldsymbol{u}_{(1)}^{k}) 
+ \Delta t \sum_{q = 1}^p S_{1q} 
\boldsymbol{f}_{(q)}^{k}.
\end{split}
\end{align}
Solution of this equation yields 
$\boldsymbol{u}_{(1)}^{k+1}$.
Notice that its right-hand side 
is determined by the known vectors
$\{\boldsymbol{u}_{(m)}^{k}: m = 1,\dots,p\}$.
Next, with the assignments \eqref{eq:assignments}
in (\ref{eq:stepdelta}), we find, upon adding
$\boldsymbol{u}_{(m+1)}^{k}$ to both sides,
\begin{align}\label{eq:corr2plus}
\begin{split}
& \boldsymbol{u}_{(m+1)}^{k+1}
- \Delta t_m
\boldsymbol{f}^I(t_{(m+1)},\boldsymbol{u}_{(m+1)}^{k+1})\\
& = \boldsymbol{u}_{(m)}^{k+1}
+ \Delta t_m
  \big(\Delta \boldsymbol{f}_{(m)}^{E,k+1}
 -\boldsymbol{f}_{(m+1)}^{I,k} \big)
\\
& + \Delta t \sum_{q = 1}^p S_{m+1,q}
\boldsymbol{f}_{(q)}^{k}
- \Delta t \sum_{q = 1}^p S_{mq}
\boldsymbol{f}_{(q)}^{k},
\end{split}
\end{align}
where we have defined the shorthand
\begin{equation}
\Delta \boldsymbol{f}_{(m)}^{E,k+1} =
\boldsymbol{f}^{E}(t_{(m)},\boldsymbol{u}_{(m)}^{k+1})
-\boldsymbol{f}^{E}(t_{(m)},\boldsymbol{u}_{(m)}^{k}).
\end{equation}
Sequential solution of this tower of equations
yields $\boldsymbol{u}_{(m+1)}^{k+1}$ for
$m = 1,\dots,p-1$.

As mentioned, for any method [Gauss-Legendre, Gauss-Radau 
(left or right), Gauss-Lobatto] Eq.~\eqref{eq:SISDCKun} defines a numerical
computed approximation to $\boldsymbol{u}(t_{n+1})$; however, note
that for Gauss-Lobatto and Gauss-Radau-right, we may also use 
$u_{(p)}^K$ for this approximation, since 
$t_{(p)} = t_n + c_p\Delta t = t_{n+1}$ for these cases.

%%%%%%%%%%%%%%%%%%%%%%%%%%%%%%%%%%%%%%%%%%%%%%%%%%%%%%%%%%%%%%%%
\section{Perturbed Kerr Initial-Data}
\label{App:ID}
%%%%%%%%%%%%%%%%%%%%%%%%%%%%%%%%%%%%%%%%%%%%%%%%%%%%%%%%%%%%%%%%

In Sec.~\ref{sec:NumericalExperiments} we use initial data
representing a nonspinning Kerr black hole with a superposed 
gravitational wave.
Initial data sets are constructed following the method
of~\cite{Pfeiffer2004}, which is based on the extended conformal thin
sandwich (XCTS) formalism. 
The Einstein constraint equations read~\cite{Pfeiffer:2005,Cook2000}
\begin{align}
\label{eq:Ham}
\SRicciS + \TrExCurv^2 - \ExCurv_{ij}\ExCurv^{ij} & = 0,\\
\label{eq:Mom}
\nabla_j\left(\ExCurv^{ij}-\SMetric^{ij}\TrExCurv\right) & = 0.
\end{align}
where  $\nabla_i$ is the covariant 
derivative compatible with the spatial metric $g_{ij}$, 
$R=g^{ij}R_{ij}$ is the trace of the Ricci tensor 
$R_{ij}$ of $g_{ij}$, and $K=g^{ij}K_{ij}$ is the 
trace of the extrinsic curvature $K_{ij}$ of the initial data hypersurface.

The conformal metric 
$\CMetric_{ij}$ and conformal factor $\CF$ are defined by 
\begin{equation}
\label{eq:SMetric-ID}
g_{ij}\equiv\CF^4\CMetric_{ij},
\end{equation}
and the time derivative of the conformal metric is denoted by
\begin{equation}
\tilde{u}_{ij}\equiv\partial_{t}\CMetric_{ij}
\end{equation}
which satisfies $\tilde{u}_{ij}\tilde{g}^{ij}=0$. 
The conformal lapse is given by $\tilde{N}=\CF^{-6}N$.
Applying this conformal 
decomposition,
Eqs.~(\ref{eq:Ham})--(\ref{eq:Mom})
can be written as 
\begin{align}
\label{eq:XCTS-Ham}
\tilde{\nabla^2}\psi-\frac{1}{8}\psi\tilde{R}-\frac{1}{12}\psi^5K^2+
\frac{1}{8}\psi^{-7}\tilde{A}_{ij}\tilde{A}^{ij} = 0,\\
\label{eq:XCTS-Mom}
\tilde{\nabla}_j\!\left(\frac{1}{2\tilde{N}}%
(\tilde{\mathbb{L}}\beta)^{ij}\right)-\tilde{\nabla}_j\!
\left(\frac{1}{2\tilde{N}}\tilde{u}^{ij}\right)
-\frac{2}{3}\psi^6\tilde{\nabla}^iK=0,
\end{align}
and the evolution equation for $K_{ij}$
yields the following equation for the lapse:
\begin{align}
\label{eq:XCTS-EvK}
\tilde{\nabla}^2(\tilde{N}\psi^7)
-\tilde{N}\psi^7\bigg(\frac{\tilde R}{8}+\frac{5}{12}\psi^4K^2
+\frac{7}{8}\psi^{-8}\tilde{A}_{ij}\tilde{A}^{ij}\bigg) \nonumber\\
=-\psi^5\left(\partial_tK-\beta^k\partial_kK\right).
\end{align}
Here $(\tilde{\mathbb{L}}\beta)^{ij}=\tilde
\nabla^i\beta^j+\tilde\nabla^j\beta^i-(2/3)\tilde
g^{ij}\tilde\nabla_k\beta^k$,
$\tilde{\nabla}_i$ is the covariant derivative compatible with 
$\tilde{g}_{ij}$, $\tilde{R}=\tilde{g}^{ij}\tilde{R}_{ij}$ is the trace of the 
Ricci tensor $\tilde{R}_{ij}$ of $\tilde{g}_{ij}$,
and $\tilde{A}^{ij}=(2\tilde{N})^{-1}
\left((\tilde{\mathbb{L}}\beta)^{ij}-
\tilde{u}^{ij}\right)$,
which is related to $K_{ij}$ by
\begin{equation}
\label{eq:K-ID}
K_{ij}=\psi^{-10}\tilde{A}_{ij}+\frac{1}{3}g_{ij}K.
\end{equation}
For given $\tilde{g}_{ij}$, $\tilde{u}_{ij}$, $K$, and $\partial_tK$,
Eqs.~\eqref{eq:XCTS-Ham},~\eqref{eq:XCTS-Mom}, and~\eqref{eq:XCTS-EvK}
are a coupled set of elliptic equations that can be solved for $\psi$,
$\tilde{N}$, and $\beta^i$. From these solutions, the physical initial
data $g_{ij}$ and $K_{ij}$ are obtained from~\eqref{eq:SMetric-ID}
and~\eqref{eq:K-ID}, respectively.

% \begin{figure}
% \includegraphics[scale=0.5]{ConstraintsID}
% \caption{
% \label{fig:ConstraintsID}
% Convergence of the elliptic solver for different amplitudes $A$. Plotted is
% the residual of the Hamiltonian constraint vs. the number of radial basis
% functions in the spherical shell containing the gravitational wave.
% }
% \end{figure}

To construct initial data describing a Kerr black hole initially in 
equilibrium, together with an ingoing pulse of gravitational waves, we make 
the following choices for the free data,
\begin{align}
\label{eq:ConformalMetric}
\tilde{g}_{ij} &= g^{\text{KS}}_{ij}+Ah_{ij},\\
\tilde{u}_{ij} &= A\partial_th_{ij}-\frac{1}{3}\tilde{g}_{ij}\tilde{g}^{kl}
A\partial_th_{kl},\\
K &= K^{KS},\\
\partial_tK &= 0.
\end{align}
In the above, $g^{\text{{KS}}}_{ij}$ and $K^{\text{KS}}$ are the
spatial metric and the trace of the extrinsic curvature in Kerr-Schild
coordinates, with mass parameter $M_{\text{KS}}=1$ and spin parameter
$a_{KS}=0$.  The pulse
of gravitational waves is denoted by $h_{ij}$ and is chosen to be an
ingoing, even parity, $m=2$, linearized quadrupole wave as given by
Teukolsky~\cite{Teukolsky1982,Rinne2008c}). The explicit expression
for the spacetime metric of the waves in spherical coordinates is
\begin{align}
h_{ij}dx^idx^j &= \left(R_1\sin^2\theta\cos2\phi\right)dr^2\notag\\
     &+ 2R_2\sin\theta\cos\theta\cos2\phi rdrd\theta\notag\\
     &- 2R_2\sin\theta\sin2\phi r\sin\theta drd\phi\notag\\
     &+ \left[R_3\left(1+\cos^2\theta\right)%
               \cos2\phi-R_1\cos2\phi\right]r^2d^2\theta\notag\\
     &+ \left[2\left(R_1-2R_3\right)\cos\theta\sin2\phi\right]r^2\sin\theta%
        d\theta d\phi\notag\\
     &+ \left[-R_3\left(1+\cos^2\theta\right)\cos2\phi%
              +R_1\cos^2\theta\cos2\phi\right]\notag\\
     &\hspace{40 mm} \times r^2\sin^2\theta d^2\phi,
\end{align}
where the radial functions are
\begin{align}
R_1 &= 3\left[\frac{F^{(2)}}{r^3}+\frac{3F^{(1)}}{r^4}+\frac{3F}{r^5}\right],\\
R_2 &= - \left[\frac{F^{(3)}}{r^2}+\frac{3F^{(2)}}{r^3}+\frac{6F^{(1)}}{r^4}%
             +\frac{6F}{r^5}\right],\\
R_3 &= \frac{1}{4}\left[\frac{F^{(4)}}{r}+\frac{2F^{(3)}}{r^2}%
             +\frac{9F^{(2)}}{r^3}+\frac{21F^{(1)}}{r^4}%
             +\frac{21F}{r^5}\right],
\end{align} 
and the shape of the waves is determined by
\begin{align}
F &= F(t+r) = F(x) = e^{-(x-x_0)^2/w^2},\\
F^{(n)} &\equiv \left[\frac{d^nF(x)}{dx^n}\right]_{x=t+r}.
\end{align}
We choose $F$ to be a Gaussian with width $w/M_{\text{KS}}=4$ 
at initial radius $x_0/M_{\text{KS}}=15$. 
The constant $A$ in Eq.~\eqref{eq:ConformalMetric} is the 
amplitude of the waves. We use the value $A=0.1$.
 
Equations~\eqref{eq:XCTS-Ham},~\eqref{eq:XCTS-Mom},
and~\eqref{eq:XCTS-EvK} are solved with the pseudospectral elliptic
solver described in~\cite{Pfeiffer2003}.

%%%%%%%%%%%%%%%%%%%%%%%%%%%%%%%%%%%%%%%%%%%%%%%%%%%%%%%%%%%%%%%%%%%%%%%%%%%%%%%
%\section*{References}
%%%%%%%%%%%%%%%%%%%%%%%%%%%%%%%%%%%%%%%%%%%%%%%%%%%%%%%%%%%%%%%%%%%%%%%%%%%%%%%

\bibliography{References/References.bib}
\end{document}